\documentclass[12pt,journal,draftclsnofoot,onecolumn]{IEEEtran}
\pdfoutput=1
\UseRawInputEncoding
\makeatletter
\def\ps@headings{%
\def\@oddhead{\mbox{}\scriptsize\rightmark \hfil \thepage}%
\def\@evenhead{\scriptsize\thepage \hfil \leftmark\mbox{}}%
\def\@oddfoot{}%
\def\@evenfoot{}}
\makeatother
\usepackage{multirow}
\usepackage{amssymb}
\usepackage{amsmath}
\usepackage{amsfonts}
\usepackage{graphicx}
\usepackage{subfigure}
\usepackage{cite}
\usepackage{enumerate}
\usepackage{url}
\usepackage{bm}
\usepackage{clrscode}
\usepackage{subfigure}
\usepackage{amsthm}
\usepackage[ruled]{algorithm2e}
\usepackage{algorithmic}
\usepackage{color}
\allowdisplaybreaks[4]
\def\oiiint{{\bigcirc}\kern-15pt{\int}\kern-9pt{\int}\kern-9pt{\int}}
\begin{document}
\title{\LARGE Intelligent Omni-Surface: Ubiquitous Wireless Transmission by Reflective-Refractive Metasurfaces}
\author{\IEEEauthorblockN{
		{Shuhang Zhang}, \IEEEmembership{Student Member, IEEE},
		{Hongliang Zhang}, \IEEEmembership{Member, IEEE},\\
		{Boya Di}, \IEEEmembership{Member, IEEE},
        {Yunhua Tan},
        {Marco Di Renzo}, \IEEEmembership{Fellow, IEEE},\\
        {Zhu Han}, \IEEEmembership{Fellow, IEEE},
        {H. Vincent Poor}, \IEEEmembership{Life Fellow, IEEE},\\
        and {Lingyang Song}, \IEEEmembership{Fellow, IEEE}
        }\\

\thanks{S. Zhang, Y. Tan, and L. Song are with Department of Electronics, Peking University, Beijing 100871 China (email: \{shuhangzhang, tanggeric, lingyang.song\}@pku.edu.cn).}
\thanks{H. Zhang and H.~V.~Poor are with the Department of Electrical Engineering, Princeton University, Princeton, NJ 08544 USA (email: hongliang.zhang92@gmail.com, poor@princeton.edu).}
\thanks{B. Di is with Department of Electronics, Peking University, Beijing 100871 China, and also with Department of Computing, Imperial College London, UK (email: diboya92@gmail.com).}
\thanks{M. Di Renzo is with Universit\'e Paris-Saclay, CNRS, CentraleSup\'elec, Laboratoire des Signaux et Syst\`emes, 91192 Gif-sur-Yvette, France. (email: marco.direnzo@centralesupelec.fr).}
\thanks{Z. Han is with Electrical and Computer Engineering Department, University of Houston, and also with the Department of Computer Science and Engineering, Kyung Hee University~(email: hanzhu22@gmail.com).}
}

\maketitle
\begin{abstract}
Intelligent reflecting surface~(IRS), which is capable to adjust propagation conditions by controlling phase shifts of the reflected waves that impinge on the surface, has been widely analyzed for enhancing the performance of wireless systems. However, the reflective properties of widely studied IRSs restrict the service coverage to only one side of the surface. In this paper, to extend the wireless coverage of communication systems, we introduce the concept of intelligent omni-surface~(IOS)-assisted communication. More precisely, IOS is an important instance of reconfigurable intelligent surface~(RIS) that is capable to provide service coverage to the mobile users~(MUs) in a reflective and a refractive manner. We consider a downlink IOS-assisted communication system, where a multi-antenna small base station (SBS) and an IOS perform beamforming jointly, to improve the received power of multiple MUs on both sides of the IOS, through different reflective/refractive channels. To maximize the sum-rate, we formulate a joint IOS phase shift design and SBS beamforming optimization problem, and propose an iterative algorithm to solve the resulting non-convex program efficiently. Both theoretical analysis and simulation results show that an IOS significantly extends the service coverage of the SBS when compared to an IRS.
\end{abstract}

\section{Introduction}
The explosive growth of the number of mobile devices has brought new user requirements and applications, and innovative networking characteristics for future communications~\cite{P2011,WZWYGW2016}, which necessitate radically novel communication paradigms~\cite{A2018}. During the past few years, there has been a growing interest in developing new transmission technologies for exploiting the implicit randomness of the propagation environment, so as to provide high-speed and seamless data services~\cite{H2019,ZZDS2019}, such as spatial modulation~\cite{YRXLH2015} and massive multiple-input and multiple-output~(MIMO) technologies~\cite{LETM2014}. However, the implementation of massive MIMO is still constrained by implementation bottlenecks, which include the hardware cost, the total energy consumption, and the high complexity for signal processing~\cite{GLHNSKL2020,LWR2020}. Due to these complexity constraints, therefore, the quality of service~(QoS) is not always guaranteed in harsh propagation environments~\cite{ABCHASZ2014}.

The recent development of metasurfaces has motivated the introduction of a new hardware technology for application to wireless communications, i.e., the reconfigurable intelligent surface~(RIS)~\cite{R2019,HKK2020}, which can improve the spectral efficiency, the energy efficiency, the security, and the communication reliability of wireless networks~\cite{R2020,LMRWTLB2020}. An RIS is an ultra-thin surface containing multiple sub-wavelength nearly-passive scattering elements~\cite{EZSSHL2020}. The sub-wavelength separation between adjacent elements of the RIS enables exotic manipulations of the signals impinging upon the surface~\cite{RZDAYRT2020,LLMHXQRA2020}. A typical implementation of an RIS consists of many passive elements that can control the electromagnetic responses of the signals through the appropriate configuration of positive intrinsic negative~(PIN) diodes distributed throughout the surface~\cite{QRLKA2020,YYHK2020}. Depending on the ON/OFF status of the PIN diodes, several signal transformations can be applied~\cite{LRLLSAQC2019}. The programmable characteristics of the RIS enables it to shape the propagation environment as desired~\cite{ZZDBHS2020}, and allows for the re-transmission of signals to the receiver at a reduced cost, size, weight, and power~\cite{HZDLSLHP2019,HHAZYZRD2020}.

In the literature, a widely studied example of RIS is referred to as intelligent reflecting surface~(IRS), in which the metasurface is designed for reflecting the signals impinging upon one side of the surface towards users located on the same side of the surface~\cite{CTY2016}. The feasibility of transmitting and receiving signals with IRSs in wireless communication systems have been verified in~\cite{AV2020} and~\cite{SAIES2021}. Based on such capabilities, research works on IRS-aided transmission include the following. In~\cite{HZADY2019}, a joint power allocation and continuous phase shift design is studied for application to a reflective IRS-assisted system in order to maximize the energy efficiency. In~\cite{ZDSH2020}, the achievable data rate of a reflective IRS-assisted communication system is evaluated and the effect of a limited number of phase shifts on the data rate is investigated. In~\cite{YXS2019}, the IRS beamforming and phase shifts design are jointly optimized to maximize the sum-rate in a reflective IRS-assisted point-to-point communication system. In~\cite{PRWXENH2020}, a reflective IRS is deployed at the cell boundary of multiple cells to assist the downlink transmission of cell-edge users, whilst mitigating the inter-cell interference.

In these research works, as mentioned, signals that impinge upon one of the two sides of a surface are completely reflected towards the same side. This implies that users located in the opposite side of the surface cannot be served by an IRS: They are out of coverage. To tackle this issue, we introduce an intelligent omni-surface~(IOS)-assisted communication system. The proposed IOS is deployed in a general multi-user downlink communication system and, in contrast with an IRS, has the dual functionality of signal reflection and refraction~\cite{DOCOMO}. More precisely, signals impinging upon one of both sides of the IOS can be simultaneously reflected and refracted towards the mobile users~(MUs) that are located on the same side and in the opposite side of the IOS, respectively~\cite{DRRT2020}. Similar to an IRS, an IOS is made of multiple passive scattering elements and programmable PIN diodes, which are appropriately designed and configured, respectively, to customize the propagation environment~\cite{ZZDTHS2020}. Unlike an IRS that completely reflects all the received signals, an IOS is capable of simultaneously reflecting and refracting the received signals~\cite{CWLZL2017}. The power ratio of the refracted and reflected signals is determined by the hardware structure of the IOS~\cite{ZZDTHPS2021}. By enabling joint reflection and refraction, an IOS provides ubiquitous wireless coverage to the MUs on both sides of it, and the propagation environment of all the users can be jointly customized by adjusting the phase shifts of the IOS scattering elements~\cite{CJS2018}. As a result, the power of the received signals can be enhanced, and the QoS of the communication links can be improved.

To serve the MUs with good performance, it is of vital importance to design the amplitude and phase response of the passive scatterers of the IOS. As mentioned in~\cite{DZSLHP2019}, the optimization of an IRS-aided system can be viewed as a joint analog beamforming design at the RIS and a digital beamforming design at the base station~(BS), so as to shape the propagation environment and improve the sum-rate of the network. In an IOS-assisted communication system, the analog beamforming performed at the IOS and the digital beamforming performed at the BS provide directional reflective/refractive radio waves to the MUs on both sides of the IOS concurrently.

However, when compared to an IRS-assisted communication system~\cite{DZSLHP2019}, the analog and digital beamforming design of an IOS-assisted communication system faces several new challenges. \emph{First}, the power of the reflected and refracted signals of the IOS may not be symmetric, i.e., the channel model of the reflected and refracted signals can be different~\cite{DRRT2020}. Therefore, the methods developed and the results obtained for IRSs may not be applicable to an IOS-assisted communication system. \emph{Second}, the power ratio of the reflected and refracted signals of an IOS can be appropriately optimized, which provides an extra degree of freedom for enhancing the communication performance~\cite{DOCOMO}. In particular, the interplay between the spatial distribution of the MUs and the optimal power ratio of the reflected and refracted signals plays an important role. \emph{Third}, the line-of-sight~(LoS) transmission link from the BS to the MUs may exist concurrently with the signals that are reflected and refracted from the IOS, which implies that the IOS needs to be optimized in order to account for the LoS links as well.

In this paper, motivated by these considerations, we aim to jointly optimize the digital beamforming at the BS and the analog reflective+refractive beamforming at the IOS, in order to maximize the sum-rate of an IOS-assisted communication system. The main contributions of this paper can be summarized as follows.
\begin{enumerate}
\item We propose a multi-user IOS-assisted downlink communication system, where an IOS is deployed to improve the QoS of several MUs that are located on both sides of the IOS. The physical characteristics of the IOS and the channel setup of the considered system model are introduced and discussed.
\item A joint BS digital beamforming and IOS analog reflective+refractive beamforming design problem is formulated to maximize the sum-rate of the MUs on both sides of the IOS. The NP-hard problem is decoupled into two subproblems: digital beamforming optimization and analog beamforming optimization, and are solved iteratively.
\item We analyse the performance of the IOS-assisted downlink communication system theoretically. The data rate gain of the MUs on different locations are derived, and the optimal power ratio of the reflected and refracted signals with respect to the spatial distribution of the MUs is solved.
\item Simulation results verify the theoretical performance of the proposed multi-user IOS-assisted downlink communication system. The numerical results also unveil that the IOS significantly improves the data rate of the MUs on both sides of it concurrently when compared to the cellular system.
\end{enumerate}

The rest of this paper is organized as follows. In Section~\ref{IOS sec}, we introduce the structure and properties of an IOS. In Section~\ref{System Model Sec}, we illustrate the considered multi-user IOS-assisted downlink communication system, including the channel model and the beamforming design. The optimization problem for the joint design of the digital beamforming at the BS and the analog reflective+refractive beamforming at the IOS is formulated in Section~\ref{Problem Formulation Sec}, and an iterative algorithm for solving the resulting non-convex problem is introduced in Section~\ref{Algorithm Sec}. The theoretical analysis of the optimal phase shifts and the optimal power ratio that maximizes the sum-rate of the network are elaborated in Section~\ref{Analysis Sec}. Numerical results are illustrated in Section~\ref{Simulation Sec} in order to quantitatively evaluate the performance of the proposed algorithm. Finally, conclusions are drawn in Section~\ref{Conclusion Sec}.

\emph{Notation:} Boldface lower and upper case symbols denote vectors and matrices, respectively. Conjugate transpose and transpose operators are denoted by $(\cdot)^{H}$ and $(\cdot)^{T}$, respectively. $\text{Tr}\{\cdot\}$ is the trace operator. $[\bm{A}]^{k,m}$ denotes the $(k,m)$th element of a matrix $\bm{A}$, and $\bm{A}^i$ is the $i$th column of matrix $\bm{A}$.
\section{Intelligent Omni-Surface}\label{IOS sec}
\begin{figure}[htbp]
\centering
\subfigure[Reflected and refracted signals from an IOS.]{
\includegraphics[width=2.7in]{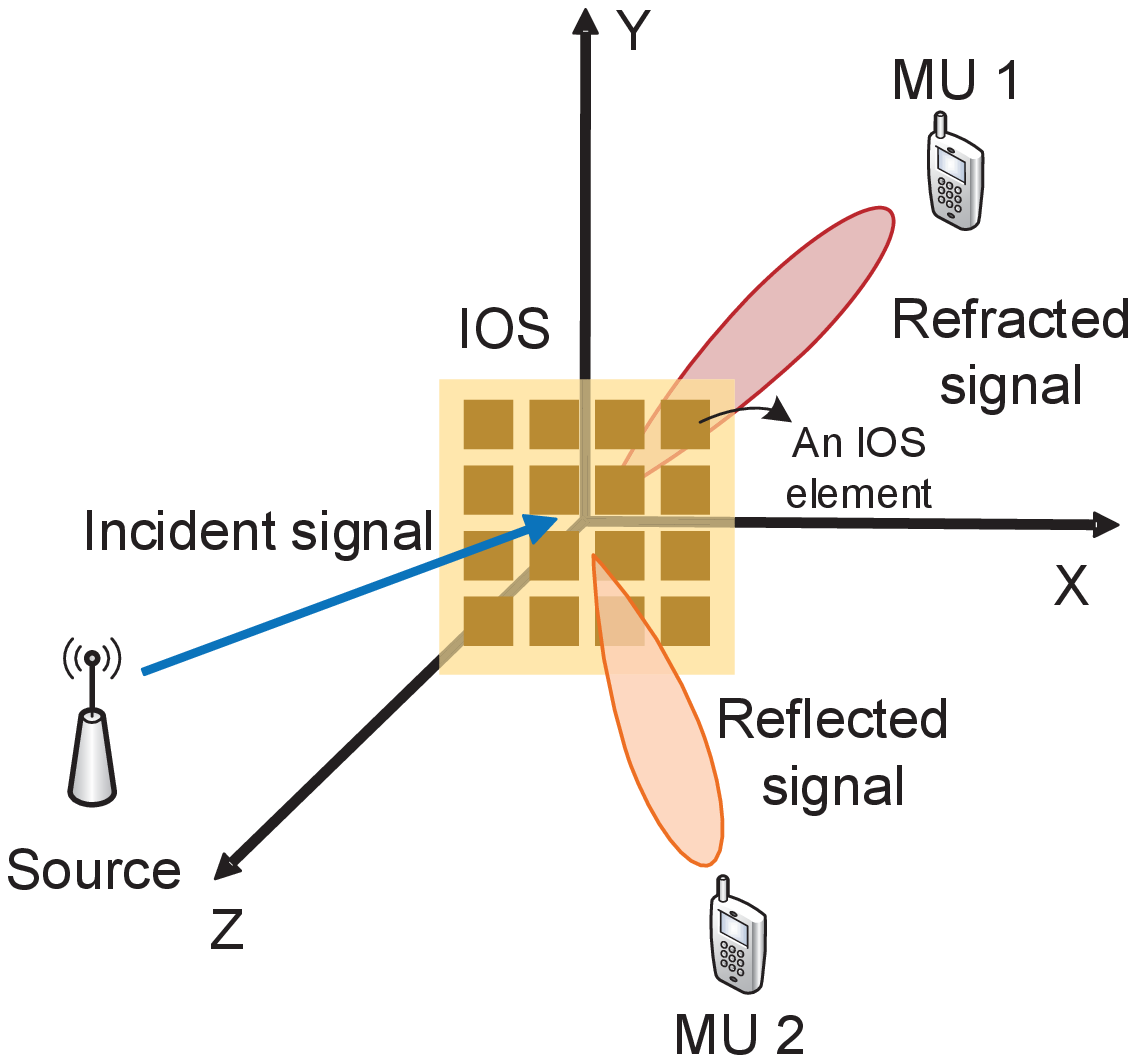}
\label{angle1}
}
\subfigure[Reflected and refracted signals for a single reconfigurable element of the IOS.]{
\includegraphics[width=2.7in]{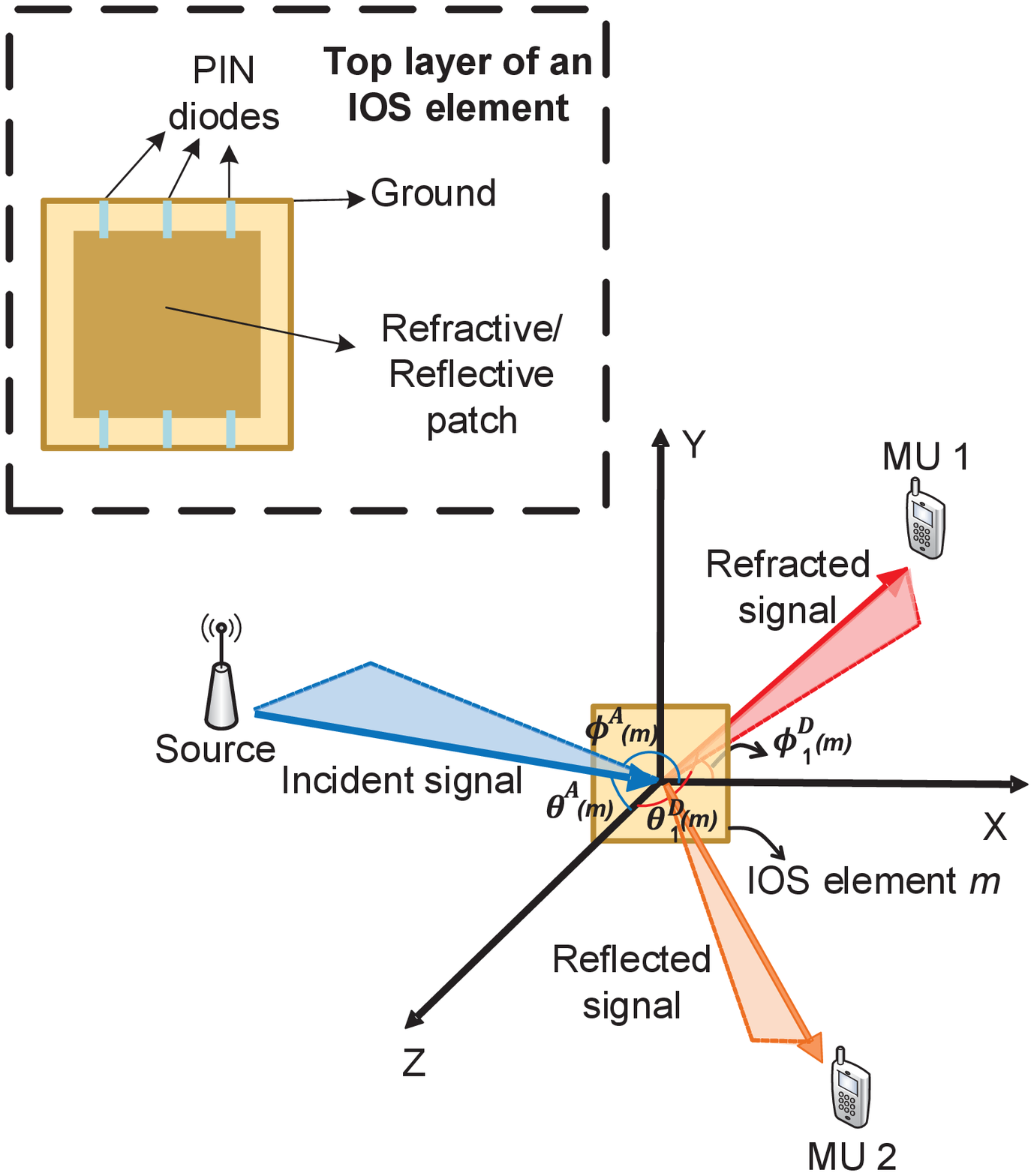}
\label{angle2}
}
\centering
\caption{Illustration of the reflected and refracted signals in an IOS-assisted communication system.}
\label{angle}
\end{figure}

An IOS is a two-dimensional array of electrically controllable scattering elements, as illustrated in Fig.~\ref{angle1}. In particular, the considered IOS is made of $M$ reconfigurable elements of equal size. The size of each element is $\delta_x$ and $\delta_y$ along the $x$ and $y$ axis, respectively. Each reconfigurable element consists of multiple metallic patches and $N_D$ PIN diodes that are evenly distributed on a dielectric substrate. The metallic patches are connected to the ground via the PIN diodes that can be switched between their ON and OFF states according to predetermined bias voltages. The ON/OFF configuration of the PIN diodes determines the phase response applied by the IOS to the incident signals. In total, each metallic patch can introduce $2^{N_D}$ different phase shifts to the incident signals. For generality, we assume that a subset of the possible phase shifts is available, which is referred to as the \emph{available phase shift set} and is denoted by $\mathcal{S}_a=\{1,\ldots,S_a\}$. The phase shift of the $m$th reconfigurable element is denoted by $\psi_m \in \mathcal{S}_a$. The $S_a$ available phase shifts are uniformly distributed with a discrete phase shift step equal to $\Delta \psi_m=\frac{2\pi}{S_a}$~\cite{ZDSH2020}. Therefore, the possible values of the phase shifts are $l_m \Delta \psi_m$, where $l_m$ is an integer satisfying $0\leq l_m \Delta \psi_m\leq S_a-1$. The vector of phase shifts of the $M$ elements of the IOS is denoted by $\bm{s}=(\phi_1,\dots,\phi_M)$. When a signal impinges, from either sides of the surface, upon one of the $M$ reconfigurable elements of the IOS, a fraction of the incident power is reflected and refracted towards the same side and the opposite side of the impinging signal. This makes an IOS different from an IRS~\cite{AB2019}. The phase shifts of the reflected and refracted signals can be either the same or different, which can be adjusted by the structure of the IOS element~\cite{P2005,I2017,RFA2018}. In this paper, we consider the case where the reflected and refracted signals of an IOS share the same phase shift.

\begin{figure}[htbp]
\centering
\subfigure[]{
\includegraphics[width=1.5in]{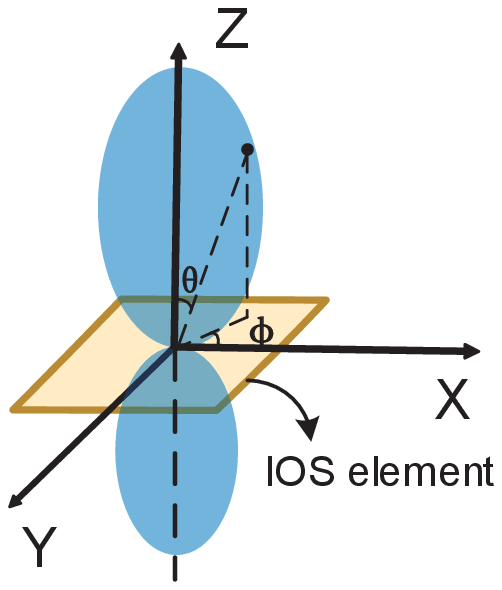}
}
\subfigure[]{
\includegraphics[width=1.5in]{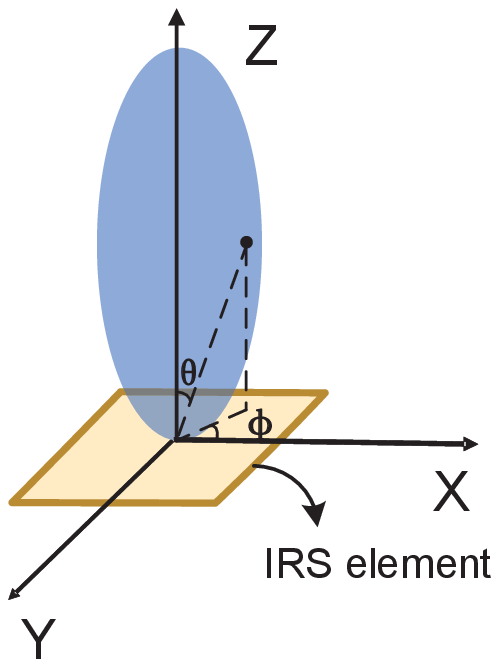}
}
\centering
\caption{Normalized power radiation pattern of (a) an IOS element, (b) an IRS element.}
\label{pattern}
\end{figure}

The direction of the signal that is emitted by the transmitter and that impinges upon the $m$th reconfigurable element of the IOS is denoted by $\xi^{A}(m)=(\theta^{A}(m), \phi^{A}(m))$, and the direction of the signal that is re-emitted by the $m$th reconfigurable element of the IOS towards the $i$th MU is denoted by $\xi_i^{D}(m)=(\theta_i^{D}(m), \phi_i^{D}(m))$, respectively, as illustrated in Fig.~\ref{angle2}. The response of the $m$th reconfigurable element of the IOS to the incident signal is denoted by the complex coefficient $g_m$, which is referred to as the \emph{amplitude gain} of the signal. In particular, $g_m$ depends on the direction of incidence i.e., $\xi^{A}(m)$, the direction of departure (either in reflection or in refraction), i.e., $\xi_i^{D}(m)$, and the phase shift $\psi_m$. In mathematical terms, we have
\begin{equation}\label{signal gain}
\begin{split}
g_m(\xi^A(m),\xi_i^D(m),\psi_m)=\sqrt{G_m K^{A}(m) K_i^{D}(m)\delta_x\delta_y|\gamma_m|^2}\exp{(-j \psi_{m})},
\end{split}
\end{equation}
where $G_m$ is the antenna power gain of the $m$th reconfigurable element, and $\psi_{m}$ is the corresponding phase shift. The coefficient $|\gamma_m|^2$ is the power ratio between the power of the signal re-emitted by the IOS and the power of the incident signal. Depending on the implementation of the IOS, $|\gamma_m|^2$ can be either a function of $\psi_m$ or a constant. In this paper, for simplicity, we assume that $|\gamma_m|^2$ is independent of the phase shift $\psi_m$~\cite{LRLLSAQC2019}. $K^{A}(m)$ and $K_i^{D}(m)$ are the normalized power radiation patterns of the incident and the re-emitted (either reflected or refracted) signal, respectively. An example for the normalized power radiation patterns is the following~\cite{TCCDHRZJCC2019}
\begin{align}\label{Ki}
K^{A}(m)=|\cos^3{\theta^{A}(m)}|,~\theta^{A}(m)\in(0,\pi),
\end{align}
\begin{align}\label{Ko}
	K_i^{D}(m)=\left\{
	\begin{aligned}
		&\frac{1}{1+\epsilon}|\cos^3{\theta_i^{D}(m)}|,~\theta_i^{D}(m)\in(0,\pi/2),\\
		&\frac{\epsilon}{1+\epsilon}|\cos^3{(\theta_i^{D}(m))}|,~\theta_i^{D}(m)\in(\pi/2,\pi),
	\end{aligned}
    \right.
\end{align}
where $\epsilon$ is a constant parameter that quantifies the power ratio between the reflected and refracted signals of the IOS, which is determined by the structure and hardware implementation of the reconfigurable elements~\cite{PG2013}. It is worth noting that the strength of both the reflected and refracted signals satisfy~(\ref{Ko}), where $\theta_i^{D}(m)\in(0,\pi/2)$ refers to the reflected signals, and $\theta_i^{D}(m)\in(\pi/2,\pi)$ refers to the refracted signals. The normalized power radiation pattern of an IOS element is illustrated in Fig.~\ref{pattern}, and it is compared against the same normalized power radiation pattern of a conventional IRS. Since the IOS is intended to be a passive device with no active power sources, the sum of the refracted and reflected power cannot exceed the power of the incident signals. Therefore, the following property holds
\begin{align}
\int_0^{2\pi}\int_0^{2\pi}|g_m(\xi^A(m),\xi^D(m),\psi_m)|^2 d\theta d\phi\leq 1.
\end{align}

\vspace{-3mm}
\section{System model}\label{System Model Sec}
In this section, we first describe the considered IOS-assisted downlink system model where a multi-antenna BS serves multiple MUs, and we then introduce the transmission channel model of the IOS-assisted system. Finally, the joint BS and IOS beamforming design for the considered transmission system is described.
\subsection{Scenario Description}
\begin{figure}[!tpb]
\centering
\includegraphics[width=5in]{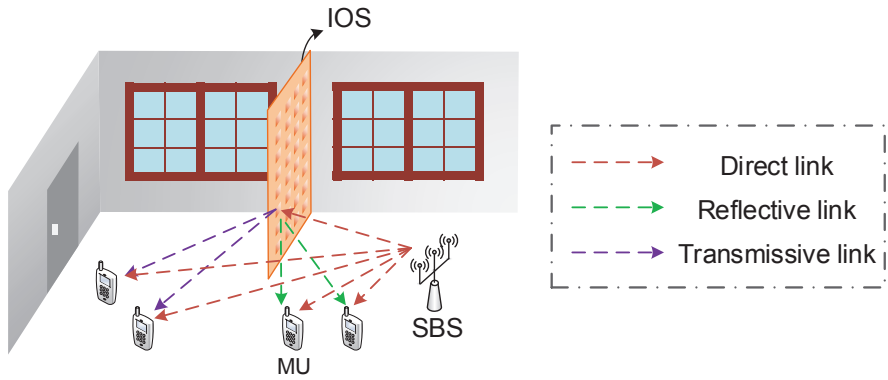}
\caption{System model for the IOS-aided downlink cellular system.}
\label{model}
\end{figure}
As shown in Fig.~\ref{model}, we consider a downlink transmission scenario in an indoor environment, which consists of one small BS~(SBS) with $K$ antennas and $N$ single-antenna MUs that are denoted by $\mathcal{N}=\{1, 2, \ldots, N\}$. Due to the complex scattering characteristics in indoor environments, some MUs that are far from the SBS may undergo severe fading, which leads to a low QoS for the corresponding communication links. To tackle this problem, we deploy an IOS in the considered indoor environment in order to extend the service coverage and to enhance the strength of the signals received by the MUs. As mentioned, the IOS consists of $M$ reconfigurable elements that are denoted by $\mathcal{M}=\{1, 2, \ldots, M\}$. The MUs are split into two subsets according to their locations with respect to the IOS. The set of MUs that receive the signals \textit{reflected} from the IOS is denoted by $\mathcal{N}_r$, and the set of MUs that receive the signals \textit{refracted} from the IOS is denoted by $\mathcal{N}_t$, with $\mathcal{N}_r\cap\mathcal{N}_t=\varnothing$, and $\mathcal{N}_r\cup\mathcal{N}_t=\mathcal{N}$. In this paper, the IOS can be viewed as an antenna array that is inherently capable of realizing analog beamforming by appropriately optimizing phase shifts of these $M$ reconfigurable elements of the IOS. This is elaborated in Section~\ref{beamforming subsection}.
\subsection{Channel Model}\label{interference sec}
The channel from the SBS to each MU consists of two parts: the reflective-refractive channel that is assisted by the IOS, and the direct path from the SBS to the MU.
\subsubsection{Reflective-Refractive Channel via the IOS}
As mentioned in Section~\ref{IOS sec}, the signal re-emitted by the IOS (see Fig.~\ref{angle1}) is given by the sum of two concurrent contributions: the refracted signal and the reflected signal. The location of each MU determines wether it receives the refracted signal or the reflected signal from the IOS. The channel from the SBS to the MU via the IOS is given by the sum of the $M$ channels through the $M$ reconfigurable elements of the IOS. Each of the $M$ SBS-IOS-MU links is model as a Rician channel in order to take into account the LoS contribution and the non-LOS~(NLoS) multipath components. In particular, the channel gain from the $k$th antenna of the SBS to the $i$th MU via the $m$th reconfigurable element of the IOS is given by
\begin{align}\label{SBS-IOS-MU channel}
h_{i,k}^m=\sqrt{\frac{\kappa}{1+\kappa}}h_{i,k}^{m,LoS}+\sqrt{\frac{1}{1+\kappa}}h_{i,k}^{m,NLoS}.
\end{align}
The LoS component of $h_{i,k}^m$ is expressed as
\begin{equation}\label{main_channel}
\begin{split}
h_{i,k}^{m,LoS}=\frac{\lambda \sqrt{G_k^{tx} K^{A}(m) G_{i}^{rx} K_i^{D}(m)} \exp\Big(\frac{-j2\pi(d_{k,m}+d_{m,i})}{\lambda}\Big)}{(4\pi)^{\frac{3}{2}}d_{k,m}^\alpha d_{m,i}^\alpha} g_m(\xi^{A}_k(m),\xi^{D}_{i}(m),\psi_m),
\end{split}
\end{equation}
where $\lambda$ is the transmission wavelength, $G_k^{tx}$ and $G_{i}^{rx}$ are the power gains of the $k$th antenna of the SBS and the antenna of the $i$th MU, respectively\footnote{The antenna of the SBS and the MUs are considered as omnidirectional, and the normalized radiation patterns of the SBS and the MUs are constants, which is not affected by the direction of the incident and received signals.}. $K^{A}(m)$ is the normalized power gain of the $k$th antenna of the SBS in the direction of the $m$th reconfigurable element of the IOS, and $K_i^{D}(m)$ is the normalized power gain of the $i$th MU in the direction of the $m$th reconfigurable element of the IOS, which are given in~(\ref{Ki}) and~(\ref{Ko}), respectively. $d_{k,m}$ and $d_{m,i}$ are the transmission distances between the $m$th reconfigurable element of the IOS and the $k$th antenna of the SBS and the $i$th MU, respectively, and $\alpha$ is the corresponding path-loss exponent. In addition, $g_m(\xi^{A}_k(m),\xi^{D}_{i}(m),\psi_m)$ is given and defined in~(\ref{signal gain}).

The NLoS component of $h_{i,k}^m$ is expressed as
\begin{equation}\label{NLoS channel gain}
h_{i,k}^{m,NLoS}=PL(k,m,i)h^{SS},
\end{equation}
where $PL(k,m,i)$ is the path-loss of the SBS-IOS-MU link given in~(\ref{main_channel}), and $h^{SS}\sim\mathcal{CN}(0,1)$ accounts for the cumulative effect of the large number of scattered paths that originate from the random scatterers available in the propagation environment.

\subsubsection{BS-MU Direct Path}
As far as the BS-MU channel is concerned, we assume a Rayleigh fading model. Therefore, the channel gain from the $k$th antenna of the SBS to the $i$th MU is
\begin{align}\label{direct channel gain}
h_{i,k}^{D}=\sqrt{G_k^{tx} F_{k,i} G_{i}^{rx} d_{i,k}^{-\alpha}}h^{SS}, \forall i \in \mathcal{N},
\end{align}
where $F_{k,i}=|\cos^3{\theta_{k,i}^{tx}}||\cos^3{\theta_{k,i}^{rx}}|$ is the normalized end-to-end power gain of the $k$th antenna of the SBS and the $i$th MU, where $\theta_{k,i}^{tx}$ is the angle between the $k$th antenna of the SBS and the direction to the $i$th MU, and $\theta_{k,i}^{rx}$ is the angle between the antenna of the $i$th MU and the direction to the $k$th antenna of the SBS. $d_{i,k}$ is the distance between the $k$th antenna of the SBS and the $i$th MU.

In summary, the channel gain from the $k$th antenna of the SBS to the $i$th MU can be written~as
\begin{align}\label{channel gain}
h_{i,k}=\sum_{m=1}^Mh_{i,k}^m+h_{i,k}^{D}, \forall i \in \mathcal{N},
\end{align}
where the first term represents the superposition of the refractive-reflective channels of the $M$ reconfigurable elements of the IOS, and the second term is the direct path.

\subsection{IOS-Based Beamforming}\label{beamforming subsection}
\begin{figure}[!tpb]
\centering
\includegraphics[width=6in]{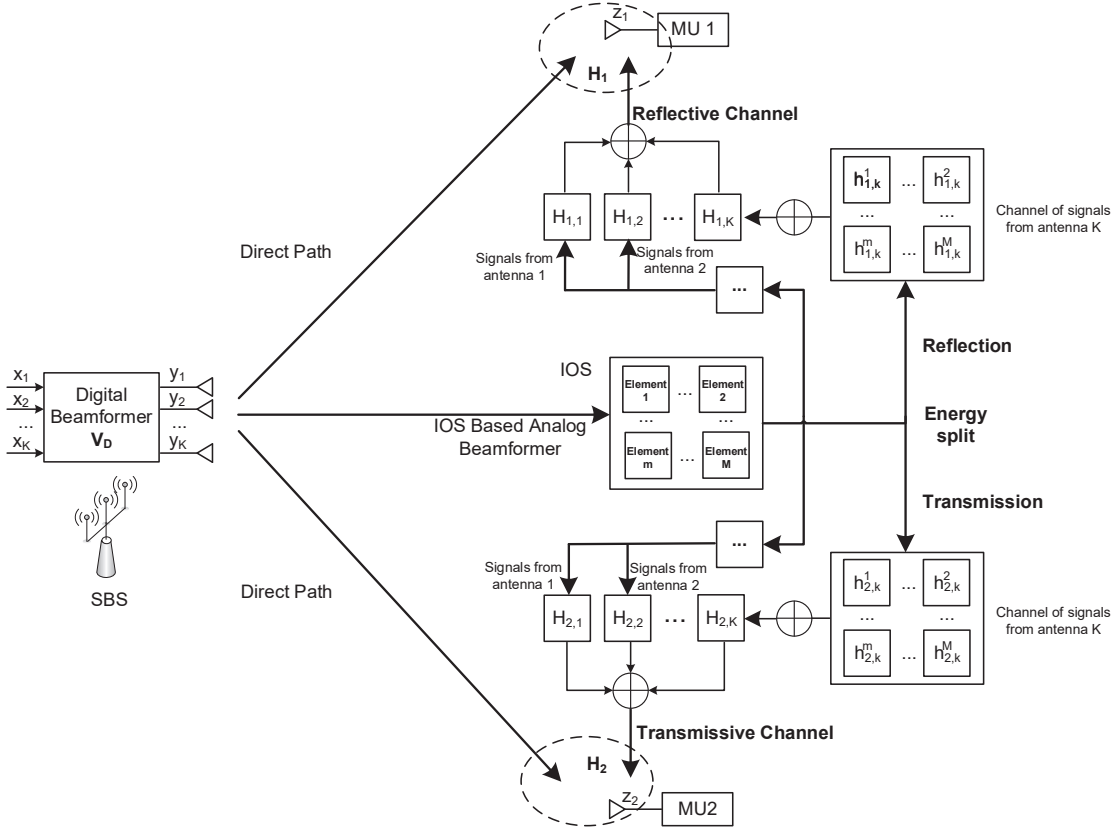}
\caption{Block diagram of the IOS-based beamforming for 2 MUs.}
\label{beamformer}
\end{figure}

In this section, we introduce the IOS-based beamforming that allows the IOS to reflect and refract the incident signals towards specified locations of the MUs. Since the reconfigurable elements of the IOS have no digital processing capabilities, we consider a hybrid beamforming scheme, where the digital beamforming is performed at the SBS and the analog beamforming is performed at the IOS. Furthermore, due to practical implementation constraints, discrete phase shifts are assumed at the IOS. An example of the hybrid beamforming for $|\mathcal{N}_r|=1$ and $|\mathcal{N}_t|=1$ is shown in Fig.~\ref{beamformer}.

\subsubsection{Digital Beamforming at the SBS}
The SBS first encodes the $N$ different data steams that are intended to the MUs via a digital beamformer, $\textbf{V}_D$, of size $K \times N$, with $K\geq N$, and then emits the resulting signals through the $K$ transmit antennas. We denote the intended signal vector for the $N$ MUs by $\textbf{x}=[x_1, x_2, \ldots, x_N]^T$. The transmitted vector of the SBS is given~by
\begin{equation}
\textbf{y}=\textbf{V}_D \textbf{x}.
\end{equation}
We denote the maximum transmission power of the SBS by $P_B$, hence the power constraint for the digital beamformer can be expressed as
\begin{equation}
\text{Tr} \left(\textbf{V}_D\textbf{V}_D^H\right)\leq P_B.
\end{equation}
\subsubsection{IOS-Based Analog Beamforming}
The received signal at the $i$th MU can be expressed as
\begin{equation}\label{zi}
z_i=\sum_{k=1}^{K} h_{i,k} \textbf{V}_D^{k,i} x_i+ \sum_{i\neq i'}\sum_{k=1}^{K} h_{i,k} \textbf{V}_D^{k,i'} x_{i'}+w_i,
\end{equation}
where $w_i$ is the additive white Gaussian noise~(AWGN) at the $i$th MU whose mean is zero and whose variance is $\sigma^2$. Therefore, the $N$ signals of the MUs in~(\ref{zi}) can be cast in a vector $\textbf{z}=[z_1, z_2, \ldots, z_N]^T$ as follows:
\begin{equation}\label{z}
\textbf{z}=\textbf{H} \textbf{V}_D \textbf{x}+\textbf{w},
\end{equation}
where $\textbf{w}=[w_1, w_2, \ldots, w_N]^T$ is the noise vector, and
$
\textbf{H}=\begin{bmatrix} h_{1,1} & \ldots & h_{1,K} \\ \ldots & \ldots  & \ldots \\ h_{N,1} & \ldots & h_{1,K} \end{bmatrix}
$
is the $N\times K$ compound channel matrix, which accounts of the propagation channel and for the phase shifts (analog beamforming) applied by the IOS (see (\ref{signal gain}) and (\ref{channel gain})).

From (\ref{z}), the downlink data rate of the $i$th MU can be formulated as
\begin{equation}\label{rate}
R_i=W_B\log_2{\left(1+\frac{|(\textbf{H}^i)^H\textbf{V}_D^i|^2}{|\sum_{i\neq i'}(\textbf{H}^{i})^H\textbf{V}_D^{i'}|^2+\sigma^2}\right)},
\end{equation}
where $W_B$ is the bandwidth. It is worth noting that~(\ref{rate}) is different from the rate in an IRS-assisted communication system, since the matrix $\textbf{H}$ accounts for both the reflective and refractive channels.
\section{Problem Formulation and Decomposition}\label{Problem Formulation Sec}
In this section, we first formulate a joint IOS phase shift design and SBS digital beamforming optimization problem to maximize the system sum-rate. Then, we decouple the resulting non-convex optimization problem into two subproblems.
\subsection{Problem Formulation}
As shown in~(\ref{main_channel}) and~(\ref{rate}), the achievable rate of an MU is determined by the phase shifts of the IOS $\bm{s}$ and by the digital beamformer of the SBS $\textbf{V}_D$. With an appropriate design of the IOS phase shifts and SBS digital beamformer, the signal-to-interference+noise ratio~(SINR) at the MUs can be improved, and the data rate of the MUs can be enhanced. To study the impact of the IOS on the MUs in terms of both reflected and refracted signals, we aim to maximize the sum-rate of these $N$ MUs in the system by jointly optimizing the IOS phase shifts $\bm{s}$ and the SBS digital beamformer $\textbf{V}_D$. In particular, this optimization problem can be formulated as
\begin{subequations}\label{formulate problem}
\begin{align}
\max_{\bm{s},\textbf{V}_D}&\sum_{i=1}^N R_i,\label{obj function}\\
\label{cons1}
s.t. & \ \text{Tr} \left(\textbf{V}_D\textbf{V}_D^H\right)\leq P_B,\\
\label{cons2}
& \  \psi_m \in \mathcal{S}_a, m=1, 2, \ldots, M.
\end{align}
\end{subequations}
In~(\ref{formulate problem}), constraint (\ref{cons1}) accounts for the maximum power budget of the SBS, and constraint (\ref{cons2}) denotes the feasible set for the phase shift of each IOS reconfigurable element.
\subsection{Problem Decomposition}
Problem (\ref{formulate problem}) is a mixed-integer non-convex optimization problem that consists of the discrete-valued variables $\bm{s}$ and the continuous-valued variables $\textbf{V}_D$. Therefore, (\ref{formulate problem}) is known to be a complex optimization problem to solve. To tackle it, we decouple (\ref{formulate problem}) into two subproblems: (i) the optimization of the digital beamforming at the SBS and (ii) the optimization of the analog beamforming (phase shifts) at the IOS.

\emph{1) Digital Beamforming Optimization at the SBS:} To optimize the digital beamforming at the SBS, we set phase shifts of the IOS, i.e. $\bm{s}$, to fixed values. Therefore, (15) reduces to
\begin{subequations}\label{digital subproblem}
\begin{align}
\max_{\textbf{V}_D}&\sum_{i=1}^N R_i,\\
s.t. & \ \text{Tr} \left(\textbf{V}_D\textbf{V}_D^H\right)\leq P_B.
\end{align}
\end{subequations}

\emph{2) Analog Beamforming Optimization at the IOS:} To optimize the phase shifts of the IOS, we assume that the digital beamforming matrix $\textbf{V}_D$ is given. Therefore, (15) reduces to
\begin{subequations}\label{analog subproblem}
\begin{align}
\max_{\bm{s}}&\sum_{i=1}^N R_i,\\
s.t. & \ \psi_m \in \mathcal{S}_a, m=1, 2, \ldots, M.
\end{align}
\end{subequations}

\section{Sum-rate Maximization: Algorithm Design}\label{Algorithm Sec}
In this section, we first design two algorithms to solve subproblems (\ref{digital subproblem}) and (\ref{analog subproblem}) individually, and we then devise an iterative algorithm for solving (\ref{formulate problem}).

\subsection{Digital Beamforming Optimization at the SBS}\label{Digital Beamforming Sec}
In this section, we solve the digital beamforming optimization at the SBS stated in~(\ref{digital subproblem}). In particular, we consider zero-forcing (ZF) beamforming and optimal transmit power optimization, in order to alleviate the interference among the MUs. As introduced in~\cite{ZDSH2020}, the ZF-based digital beamforming can be formulated as
\begin{equation}\label{beam-signal}
\textbf{V}_D=\textbf{H}^H(\textbf{H}\textbf{H}^H)^{-1}\textbf{P}^{1/2}=\widetilde{\textbf{V}}_D\textbf{P}^{1/2},
\end{equation}
where $\widetilde{\textbf{V}}_D=\textbf{H}^H(\textbf{H}\textbf{H}^H)^{-1}$ and $\textbf{P}$ is a diagonal matrix whose $i$th diagonal element, which is denoted by $p_i$, is the received power at the $i$th MU.

Based on the ZF beamforming design in (\ref{beam-signal}), problem~(\ref{digital subproblem}) can be simplified as
\begin{subequations}\label{simplified digital subproblem}
\begin{align}
\max_{p_i\geq0}&\sum_{i=1}^N W_B\log_2\left(1+\frac{p_i}{\sigma^2}\right),\\
s.t. & \ \text{Tr} \left(\textbf{P}^{1/2}\widetilde{\textbf{V}}_D^H\widetilde{\textbf{V}}_D\textbf{P}^{1/2}\right)\leq P_B.
\end{align}
\end{subequations}
The optimal solution of problem (\ref{simplified digital subproblem}) is the well-known water-filling power allocation~\cite{TV2005}
\begin{equation}
p_i=\frac{1}{\nu_i}\max\left(\frac{1}{\mu}-\nu_i\sigma^2, 0\right),
\end{equation}
where $\nu_i$ is the $i$th diagonal element of $\widetilde{\textbf{V}}_D^H\widetilde{\textbf{V}}_D$ and $\mu$ is a normalization factor that fulfills the constraint $\sum_{i=1}^N \max\left(\frac{1}{\mu}-\nu_i\sigma^2, 0\right)=P_B$. After obtaining $\textbf{P}$ through the water-filling algorithm, the digital beamforming matrix is directly obtained from~(\ref{beam-signal}).
\subsection{Analog Beamforming Optimization at the IOS}\label{IOS Phase Shift Design Sec}
In this section, we solve the analog beamforming optimization at the IOS stated in~(\ref{analog subproblem}). Since the digital beamforming is assumed to be fixed and to be given by the ZF-based precoding matrix in (\ref{simplified digital subproblem}), the problem in (\ref{analog subproblem}) can be simplified as
\begin{subequations}\label{reformulate analog subproblem}
\begin{align}
\max_{\bm{s}}&\sum_{i=1}^N W_B\log_2{\left(1+\frac{|(\textbf{H}^i)^H\textbf{V}_D^i|^2}{\sigma^2}\right)}, \label{reformulate obj}\\
s.t. & \ 0\leq \psi_m<2\pi, m=1, 2, \ldots, M.
\end{align}
\end{subequations}

Problem (\ref{reformulate analog subproblem}) is tackled in two steps: (i) first, a relaxed problem that assumes continuous phase shifts is considered, and (ii) then, a branch-and-bound based algorithm is proposed to account for the discrete phase shifts.
\subsubsection{Continuous IOS Phase Shift Design}\label{Continuous IOS Phase Shift Design Sec}
According to the channel model introduced in Section~\ref{interference sec}, the objective function in~(\ref{reformulate obj}) can be approximated to a concave function with respect to the phase shift of each reconfigurable element of the IOS. Therefore, problem~(\ref{reformulate analog subproblem}) can be solved by optimizing phase shifts of these $M$ reconfigurable elements iteratively. More precisely, the downlink data rate of each MU, i.e., $R_i$, can be approximated to a concave function of each variable in $\bm{s}$. Therefore, the objective function in~(\ref{reformulate obj}), i.e., $\sum_{i=1}^N R_i$, is approximately a concave function of each variable in $\bm{s}$ while keeping the others fixed. To optimize the phase shifts of all the reconfigurable elements of the IOS, we first set a random initial solution, which is denoted by $\bm{s}^0=(s_1^0,\dots,s_M^0)$. Then, we iteratively optimize the phase shift of each reconfigurable element. Without loss of generality, let us consider the optimization of $\psi_m$ at the $r$th iteration. We fix the phase shift of all the other reconfigurable elements at the newly solved value, i.e., $s_1=s_1^r, \ldots, s_{m-1}=s_{m-1}^r, s_{m+1}=s_{m+1}^{r-1}, \ldots, s_M=s_M^{r-1}$, and maximize $\sum_{i=1}^N W_B\log_2{\left(1+\frac{|(\textbf{H}^i)^H\textbf{V}_D^i|^2}{\sigma^2}\right)}$ as a function of only $\psi_m$ as mentioned above. The obtained solution $\psi_m^r$ is then updated as a temporary solution for the phase shift of the $m$th reconfigurable element, and the corresponding sum-rate increment is given by
\begin{equation}\label{deltaR}
\Delta R_m^r=\sum_{i=1}^N \log_2{\left(1+\frac{|(\textbf{H}^i)^H\textbf{V}_D^i|^2}{\sigma^2}\right)}|_{s_{m}^r,}-\sum_{i=1}^N \log_2{\left(1+\frac{|(\textbf{H}^i)^H\textbf{V}_D^i|^2}{\sigma^2}\right)}|_{s_{m}^{r-1}}.
\end{equation}
The algorithm terminates when the increment of the downlink sum-rate between two consecutive iterations is below a specified threshold, i.e., $\sum_{m=1}^M \Delta R_m^r<R_{th}$. Since problem~(\ref{reformulate analog subproblem}) is non-convex with respect to all the variables in $\bm{s}$ jointly, the proposed solution converges to a locally optimal solution. The convergence of the proposed algorithm is proved in Section~\ref{Algorithm-subsec}. The proposed continuous IOS phase shift design is summarized in Algorithm~\ref{continuous phase shift algorithm}.

\begin{algorithm}[t]
\caption{Continuous IOS Phase Shift Design.}
\begin{algorithmic}[1]\label{continuous phase shift algorithm}
\STATE {\textbf{Initialization:} Set an initial solution $\bm{s}^0=(s_1^0,\dots,s_M^0)$ to problem~(\ref{reformulate analog subproblem})}
\STATE {\textbf{While} $\sum_{m=1}^M \Delta R_m^r\geq R_{th}$}
\STATE {\quad \textbf{For} $m=1:M$}
\STATE {\quad\quad Solve $\psi_m$ in problem (\ref{reformulate analog subproblem}) while keeping the other variables fixed}
\STATE {\quad\quad Compute $\Delta R_m^r$ in~(\ref{deltaR})}
\STATE {\quad\quad Update the IOS phase shifts to $\bm{s}=(s_1=s_i^r, \cdots, s_{m}=s_{m}^r, s_{m+1}=s_{m+1}^{r-1}, \cdots, s_M=s_M^{r-1})$}
\STATE {\textbf{Output:} $\bm{s}$}
\end{algorithmic}
\end{algorithm}

\subsubsection{Discrete IOS Phase Shift Design}
By applying Algorithm~\ref{continuous phase shift algorithm}, the continuous phase shifts of the $M$ reconfigurable elements are obtained, which are denoted by $\psi_m^{opt}, m=1, 2, \ldots, M$. However, $\psi_m^{opt}$ may not correspond to any of the finite and discrete phase shifts in $\mathcal{S}_a$. In general, in fact, the continuous phase shift of the $m$th reconfigurable element of the IOS that is solution of problem (\ref{reformulate analog subproblem}) lies in the range determined by the two consecutive phase shifts equal to $l_m \Delta \phi_m$ and $(l_m+1) \Delta \phi_m$, i.e., $l_m \Delta \phi_m \leq \psi_m^{opt}\leq (l_m+1) \Delta \phi_m$. Therefore, after computing the continuous phase shifts of the IOS with the aid of Algorithm~\ref{continuous phase shift algorithm}, the search space for the $M$ discrete phase shifts still encompasses $2^M$ possibilities. To overcome the associated computational complexity, we propose an efficient branch-and-bound algorithm that yields the optimal discrete phase shifts of the IOS that belong to the finite set $\mathcal{S}_a$.

\begin{figure}[!tpb]
\centering
\includegraphics[width=4in]{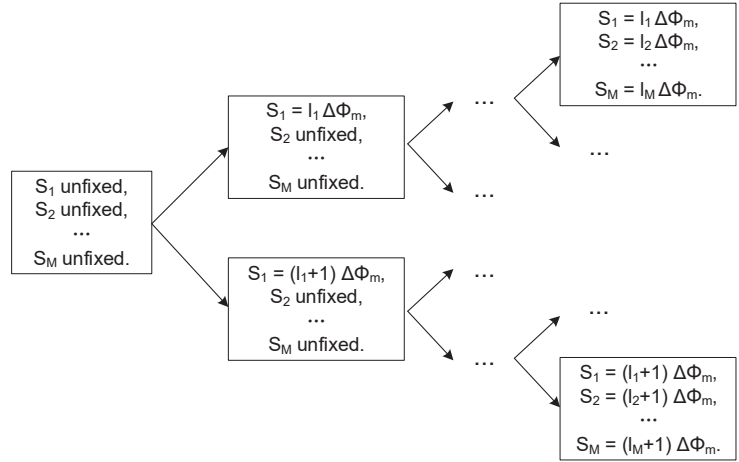}
\caption{Tree structure of the proposed branch-and-bound algorithm.}
\label{tree}
\end{figure}

The solution space of the IOS phase shift vector~$\bm{s}$ can be viewed as a binary tree structure, as shown in Fig.~\ref{tree}. Each node of the tree contains the phase shift information of all the $M$ reconfigurable elements, i.e., $\bm{s}=(s_1,\dots,s_M)$. At the root node, all the variables in~$\bm{s}$ are unfixed. After applying Algorithm~\ref{continuous phase shift algorithm}, the value of an unfixed variable $\psi_m$ at a father node can be one of the two phase shifts $l_m \Delta \phi_m$ or $(l_m+1) \Delta \phi_m$. As illustrated in Fig.~\ref{tree}, this branches the father node into two child nodes. This operation can be repeated for each parent node. Our objective is to devise an efficient algorithm that allows us to solve problem~(\ref{analog subproblem}) based on the tree structure illustrated in Fig.~\ref{tree}, which is determined by Algorithm~\ref{continuous phase shift algorithm}.

The proposed algorithm starts by randomly setting a feasible solution for the phase shift $\psi_m$ to any of the two possible solutions $l_m \Delta \phi_m$ or $(l_m+1) \Delta \phi_m$. The corresponding value of the objective function yields a lower bound to problem~(\ref{reformulate analog subproblem}). Then, the proposed branch-and-bound algorithm recursively splits the search space of problem~(\ref{analog subproblem}) into smaller spaces based on the tree structure illustrated in Fig.~\ref{tree}. The search is started from the root node of the tree in a depth-first-search method. At each node, one of the unfixed variable is fixed, which branches the node into two child nodes. An upper bound for the sum-rate of each of the two child nodes is obtained solving problem~(\ref{reformulate analog subproblem}) by relaxing the unfixed discrete variables $\bf{s}^{uf}$ into continuous variables. The obtained upper bound for the rate is compared with the current lower bound, and, depending on their relation, the branch that corresponds to the node under analysis is either kept or pruned. The lower bound is compared to the sum-rate of a node if all its variables are fixed, and is updated to the larger one between the two values. The process is iterated until all the nodes of the tree in Fig.~\ref{tree} are either visited or pruned. The details of the branch-and-bound method that solves the IOS phase shift design problem~(\ref{analog subproblem}) is summarized in Algorithm~\ref{IOS Phase Shift Design}.

\begin{algorithm}[t]
\caption{IOS Phase Shift Design.}
\begin{algorithmic}[1]\label{IOS Phase Shift Design}
\STATE {\textbf{Initialization:} Set an initial feasible solution $\bm{s}$ to problem~(\ref{analog subproblem}) based on Algorithm~\ref{continuous phase shift algorithm} and set the corresponding sum-rate $R_{lb}$ as the lower bound}
\STATE {\textbf{While} All nodes in Fig.~\ref{tree} are not visited or pruned}
\STATE {\quad Calculate the upper bound of the current node~$R_{ub}=\max_{\bm{s}^{uf}}\sum_{i=1}^N \log_2{\left(1+\frac{|(\textbf{H}^i)^H\textbf{V}_D^i|^2}{\sigma^2}\right)}$}
\STATE {\quad \textbf{If} $R_{ub}<R_{lb}$: Prune the corresponding branch; Return}
\STATE {\quad \textbf{Else}:}
\STATE {\quad\quad \textbf{If} The current node has any child nodes: Move to one of its two child nodes; Continue}
\STATE {\quad\quad \textbf{Else} Calculate the corresponding sum-rate $R_{curr}$}
\STATE {\quad\quad \textbf{If} $R_{curr}>R_{lb}$: $R_{lb}=R_{curr}$; Return}
\STATE {\textbf{Output:}$\bm{s}$}
\end{algorithmic}
\end{algorithm}
\subsection{Joint SBS Digital Beamforming and IOS Phase Shift Optimization}\label{Algorithm-subsec}
Based on the proposed solutions for the ZF-based digital beamforming at the SBS and the branch-and-bound algorithm for obtaining the discrete phase shifts of the IOS by leveraging Algorithm~\ref{continuous phase shift algorithm} and Algorithm~\ref{IOS Phase Shift Design}, problem (\ref{formulate problem}) can be solved by using alternating optimization as summarized in Algorithm~\ref{Joint Algorithm}. Given the phase shifts of the reconfigurable elements, the analog beamforming can be considered as part of the channel vector, and ZF precoding of the SBS digital beamforming can be designed accordingly to eliminate the interference~\cite{RPLLMET2013}. When the phase shifts of the reconfigurable elements are changed, the ZF precoding should be adjusted correspondingly. The convergence and complexity of Algorithm~\ref{Joint Algorithm} are analyzed in the following two propositions.

\begin{algorithm}[t]
\caption{Joint SBS Digital Beamforming and IOS Phase Shift Optimization.}
\begin{algorithmic}[1]\label{Joint Algorithm}
\STATE {\textbf{While} The sum-rate difference between two consecutive iterations is below a threshold $\omega$}
\STATE {\quad Perform SBS digital beamforming as introduced in Section~\ref{Digital Beamforming Sec}}
\STATE {\quad Compute the continuous IOS phase shifts by leveraging Algorithm~\ref{continuous phase shift algorithm}}
\STATE {\quad Compute the discrete IOS phase shifts by leveraging Algorithm~\ref{IOS Phase Shift Design}}
\STATE {\quad Update the ZF precoding of the SBS digital beamforming}
\STATE {Obtain the maximum sun-rate of the IOS-assisted communication system}
\end{algorithmic}
\end{algorithm}

\textbf{Proposition 1:} The proposed joint SBS digital beamforming and IOS analog beamforming optimization algorithm is convergent to a local optimum solution.
\begin{proof}
We denote the sum-rate at the $r$th iteration by $\mathcal{R}(\bm{s}^{r},\textbf{V}_D^{r})$. At the $(r+1)$th iteration, the solution of the optimal SBS digital beamforming given $\bm{s}=\bm{s}^{r}$ yields a sum-rate $\mathcal{R}(\bm{s}^{r},\textbf{V}_D^{r+1})\geq\mathcal{R}(\bm{s}^{r},\textbf{V}_D^{r})$. Similarly, the solution of the optimal IOS analog beamforming given $\textbf{V}_D=\textbf{V}_D^{r+1}$ yields a sum-rate $\mathcal{R}(\bm{s}^{r+1},\textbf{V}_D^{r+1})\geq\mathcal{R}(\bm{s}^{r},\textbf{V}_D^{r+1})$. Therefore, we obtain the following inequalities $\mathcal{R}(\bm{s}^{r+1},\textbf{V}_D^{r+1})\geq\mathcal{R}(\bm{s}^{r},\textbf{V}_D^{r+1})\geq\mathcal{R}(\bm{s}^{r},\textbf{V}_D^{r})$, i.e., the sum-rate does not decrease at each iteration of Algorithm~\ref{Joint Algorithm}. Since the sum-rate is upper bounded thanks to the constraint on the total transmit power, Algorithm~\ref{Joint Algorithm} converges in a finite number of iterations to a local optimum solution.
\end{proof}

\textbf{Proposition 2:} The complexity of each iteration of the proposed joint SBS digital beamforming and IOS analog beamforming optimization algorithm is $O(2^M+N^2)$.
\begin{proof}
At each iteration, the SBS digital beamforming problem for the $N$ MUs can be solved by using convex optimization methods. As mentioned in~\cite{B2014}, the minimum complexity of convex optimization is $O(N)$. The complexity of the IOS analog beamforming problem is the sum of the complexities of Algorithm~\ref{continuous phase shift algorithm} and Algorithm~\ref{IOS Phase Shift Design}. Algorithm~\ref{continuous phase shift algorithm} is also based on convex optimization, with the complexity being $O(N)$, and the complexity of the branch-and-bound based Algorithm~\ref{IOS Phase Shift Design} is, in the worst case, $O(2^M)$. Therefore, the complexity of each iteration of Algorithm~\ref{Joint Algorithm} is $O(N)+O(N)+O(2^M)=O(2^M+N)$.
\end{proof}
\section{Performance Analysis of the IOS-Assisted Communication System}\label{Analysis Sec}
In this section, we first discuss the impact of the phase shift design on the reflected and refracted signals, and we then analyze the downlink sum-rate as a function of the power ratio of the reflected and refracted signals. In this section, for ease of understanding, the impact of small-scale fading is ignored, and only the impact of the distances, the radiation pattern of the reconfigurable elements of the IOS, and the power allocation ratio between the reflected and refracted signals are considered.
\subsection{Analysis of the Phase Shift Design}
Our objective is to study the relation between the optimal phase shifts of the IOS when it is used as a refractive and reflective surface, and to understand the differences between these two cases. For analytical convenience, we assume that the power ratio $|\gamma_m|^2$ between the power of the signal re-emitted by the IOS and the power of the incident signal is a constant value.



In order to understand the differences and similarities between an IRS (only reflections are allowed) and an IOS (both reflections and refractions are allowed simultaneously), the following proposition considers the case study with two MUs, when one MU (the $i$th MU) lies in the reflective side of the IOS and the other MU (the $j$th MU) lies in the refractive side of the IOS.

\begin{figure}[t]
    \centering
    \includegraphics[width=4.5in]{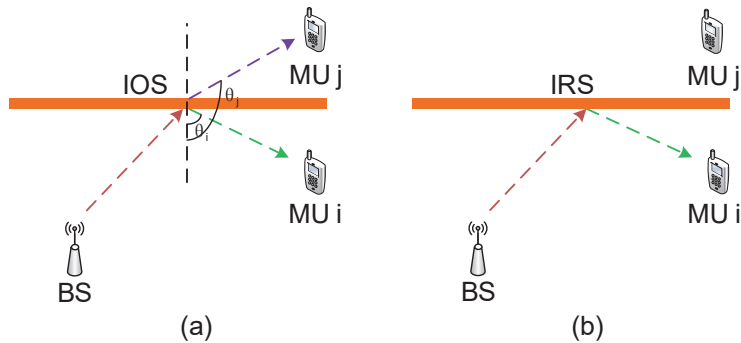}
    \caption{Examples of MU distributions for the performance analysis of an IOS-assisted communication system.  (a) IOS setup in which two MUs are located symmetrically with respect to the IOS; (b) IRS setup in which only reflections are allowed.}
    \label{appendix D}
\end{figure}

\textbf{Proposition 3:} Consider an IOS-assisted communication system with two MUs, which are denoted by the indices $i$ and $j$. If the two users are located symmetrically with respect to the IOS, as shown in Fig.~\ref{appendix D}(a), and the impact of small-scale fading is ignored, the optimal SBS digital beamforming and IOS analog beamforming are the same for both MUs.
\begin{proof}
See Appendix A.
\end{proof}

In an IOS-assisted communication system, IOS phase shifts are designed for the MUs on both sides of the IOS jointly. The optimal IOS phase shift design for the MUs on both sides of the IOS may be different from that only consider the MUs in $\mathcal{N}_r$ (i.e., in the reflective side of the surface) or $\mathcal{N}_t$ (i.e., in the refractive side of the surface). Proposition 3, in particular, can be regarded as a special case in which the optimal solution can be guaranteed for the two users in both sides of the IOS. To maximize the downlink sum-rate for multiple MUs located in both sides of the IOS, we introduce a priority-based approach that gives higher priority to either the MUs on the reflective or refractive sides of the IOS. In particular, the \emph{priority index} of the $i$th MU that describes the strength of the reflective/refractive channel gain with respect to its location is defined as
\begin{align}\label{priority index}
	\mathcal{P}_i=\left\{
	\begin{aligned}
		&\sum_{m=1}^M \frac{|\cos^3{\theta_i^{D}(m)}|}{(1+\epsilon)d_{m,i}^\alpha},~i \in \mathcal{N}_r,\\
		&\sum_{m=1}^M \frac{\epsilon|\cos^3{\theta_i^{D}(m)}|}{(1+\epsilon)d_{m,i}^\alpha},~i \in \mathcal{N}_t
	\end{aligned}
    \right.
\end{align}

\textbf{Proposition 4:} Let us assume that the distance between the SBS and the IOS is fixed and only the distances of the MUs change, as well as that the small-scale fading is not considered. Thanks to the IOS, the MU with the highest priority index in~(\ref{priority index}) obtains the largest data rate gain.
\begin{proof}
See Appendix B.
\end{proof}

Based on Proposition 4, the following remarks for some asymptotic regimes of the IOS-assisted communication system can be made.

\textbf{Remark 1:} When $\epsilon\rightarrow 0$, the IOS boils down to an IRS, and $\sum_{i\in \mathcal{N}_r}\mathcal{P}_i\gg\sum_{i\in \mathcal{N}_t}\mathcal{P}_i$ is satisfied. The IOS phase shift design only considers the MUs that belong to the set $\mathcal{N}_r$.

\textbf{Remark 2:} When $\epsilon\rightarrow \infty$, the IOS only refracts the signals to the opposite side of the SBS, and $\sum_{i\in \mathcal{N}_r}\mathcal{P}_i\ll\sum_{i\in \mathcal{N}_t}\mathcal{P}_i$ is satisfied. The IOS phase shift design only considers the MUs that belong to the set $\mathcal{N}_t$.

\subsection{Analysis of the Refraction/Reflection Power Ratio}
In this section, we analyze the impact of the power ratio of the reflected and refracted signals $\epsilon$ on the sum-rate of an IOS-assisted communication system. 

\textbf{Proposition 5:} Given the average distance between the IOS and the MUs, the power ratio of the reflected and refracted signals $\epsilon$ is positively correlated with the ratio of the number of MUs on the two sides of the IOS, i.e., $\epsilon \propto \mathcal{N}_t/\mathcal{N}_r$.
\begin{proof}
See Appendix C.
\end{proof}

\textbf{Proposition 6:} Given a pair of symmetrically located MUs as illustrated in Fig.~\ref{appendix D}(a), a larger proportion of the available power is allocated to the MUs with a weak direct link (i.e., low received power). More specifically, $\epsilon > 1$ when the sum of the received power of the direct links from the SBS to the MUs in $\mathcal{N}_r$ is larger than that of the MUs in $\mathcal{N}_t$. Otherwise, $\epsilon<1$ holds.
\begin{proof}
See Appendix D.
\end{proof}

In particular, when the distance from the SBS to the IOS is much larger than the distance from the IOS to the MUs, the distance from the SBS to each MU is approximately the same. Therefore, (\ref{appendixC equation}) in Appendix C can be simplified as
\begin{equation}\label{24}
\frac{\emph{d} (R_i+R_j)}{\emph{d}\epsilon}\simeq \frac{1}{\ln2}\frac{\beta_i^2(1-\epsilon)(1+\epsilon)}{(\epsilon\beta_i(1+\epsilon)+(1+\epsilon)^2)(\beta_i(1+\epsilon)+(1+\epsilon)^2)}.
\end{equation}
From (\ref{24}), we evince that the maximum value of $\sum_{i \in \mathcal{N}}\Delta R_i$ is obtained for $\epsilon=1$. Therefore, the following remark follows.

\textbf{Remark 3:} When the distance from the SBS to the IOS is much larger than the distance from the IOS to the MUs, the IOS maximizes the throughput of the system for $\epsilon=1$, i.e., the power of the refracted and reflected signals is the same.

The assumptions in Proposition 6 and Remark 3 are usually satisfied when the IOS is deployed at the cell edge for coverage extension. Therefore, an IOS with $\epsilon=1$ is capable of maximizing the sum-rate of the MUs at the cell edge of an IOS-assisted communication system. Finally, the following proposition yields the largest theoretical gain that an IOS-assisted system provides with respect to the benchmark IRS-assisted system.

\textbf{Proposition 7:} The ratio of the downlink sum-rate of an IOS-assisted system and an IRS-assisted system is upper-bounded by two.
\begin{proof}
See Appendix E.
\end{proof}
Proposition 7 unveils that an IOS may double the downlink sum-rate when compared to an IRS. This upper-bound is, however, difficult to be attained because it requires that the signal-to-interference-plus-noise ratio is infinite, as discussed in Appendix E.

\section{Simulation Results}\label{Simulation Sec}
In this section, we evaluate the performance of the considered IOS-assisted system based on the proposed algorithm, and compare it with and IRS-assisted system~\cite{DZLSLH2020} and a conventional cellular system in the absence of IRS or IOS. In the IRS-assisted system, the IRS only reflects the signals from the SBS to the MUs, and the surface does not work in refraction mode. In the conventional cellular system, the MUs receive only the direct links from the SBS without the assistance of a reconfigurable surface.

In the simulations, we set the height of the SBS and the center of the IOS at 2 m, and the distance between the SBS and the IOS is 100 m. The MUs are randomly deployed within a disk of radius 50 m centered at the IOS. The maximum transmit power of the SBS is $P_B=40$ dBm, the carrier frequency is 5.9 GHz, the noise power is -96 dBm, the antenna separation at the SBS is 0.2 m, and the inter-distance between the reconfigurable elements of the IOS is 0.025 m (i.e., half of the wavelength). The numbers of MUs ($N$) and SBS antennas ($K$) are 5, and the power ratio of the reflected and refracted signals is $\epsilon=1$ (unless stated otherwise). The path-loss exponent of the direct link is 3, and the Rician factor is $\kappa=4$.

\begin{figure}[htbp]
\centering
\subfigure[Number of IOS reconfigurable elements~($\sqrt{M}$) vs. sum-rate ($S_a=4$).]{
\includegraphics[width=3in]{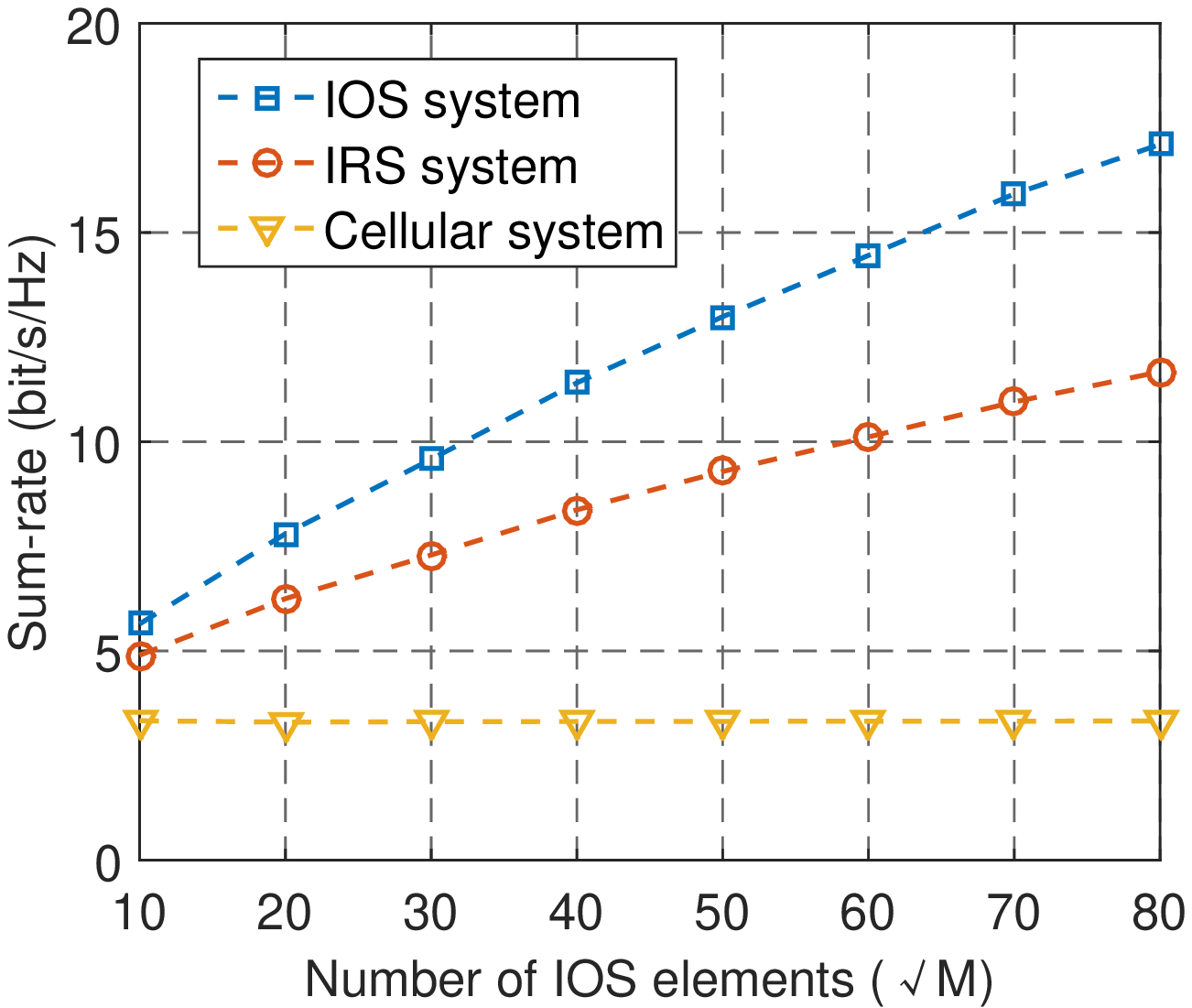}
\label{simulation1-1}
}%
\subfigure[Quantization bits of each IOS element~($N_D$) vs. sum-rate.]{
\includegraphics[width=3in]{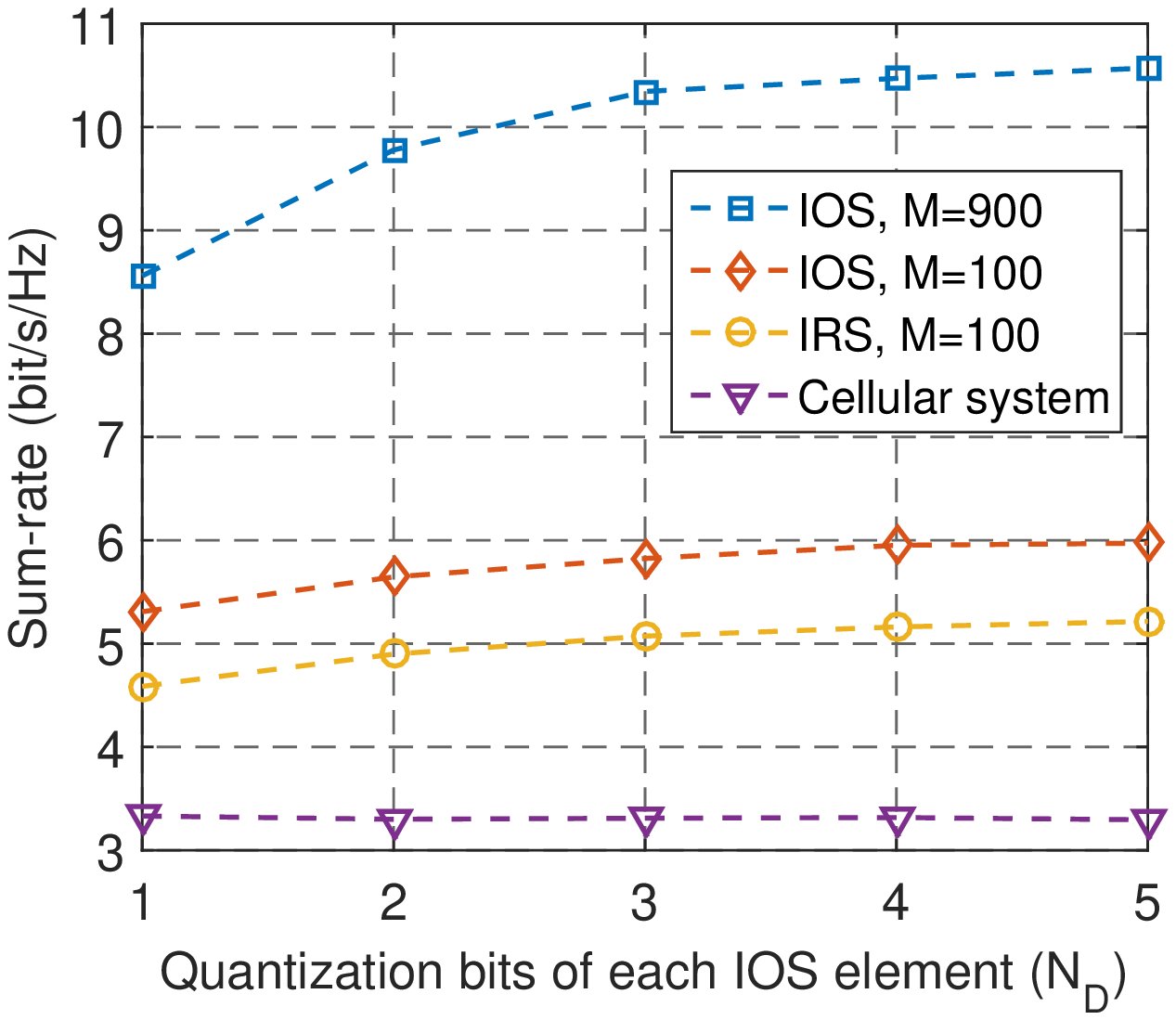}
\label{simulation1-2}
}%
\centering
\caption{Impact of the surface size and quantization bits on the sum-rate.}
\label{simulation1}
\end{figure}


Fig.~\ref{simulation1-1} illustrates the average sum-rate of the MUs as a function of the size of the IOS. The average sum-rate is defined as the average of the sum-rate in (\ref{obj function}) as a function of the spatial distribution of the MUs. The IOS is modeled as a square array with $\sqrt{M}$ elements on each line and each row. The average sum-rate increases with the number of IOS reconfigurable elements, and the growth rate gradually decreases with the IOS size. An IOS with $30\times30$ elements improves the average sum-rate of about 2.9 times when compared to a conventional cellular system. On the other hand, an IRS of the same size improves the average sum-rate of about 2.3 times only. The IOS provides a higher average sum-rate since it is capable of supporting the refraction of the MUs that are located on both sides of the surface. The rate improvement offered by an IOS over an IRS is less than two, which is in agreement with Proposition 7.


Fig.~\ref{simulation1-2} shows the average sum-rate as a function of the number of quantization bits $N_D$ of each reconfigurable element of the IOS, with $N_D=\log_2(S_a)$. The average sum-rate increases with $N_D$ and converges to a stable value as $N_D$ increases. We observe, in particular, that few quantization bits are sufficient to achieve most of the average sum-rate and that the convergence rate increases with the size $M$ of the IOS.

\begin{figure}[htbp]
\centering
\subfigure[MU distribution radius: 2 m.]{
\includegraphics[width=3in]{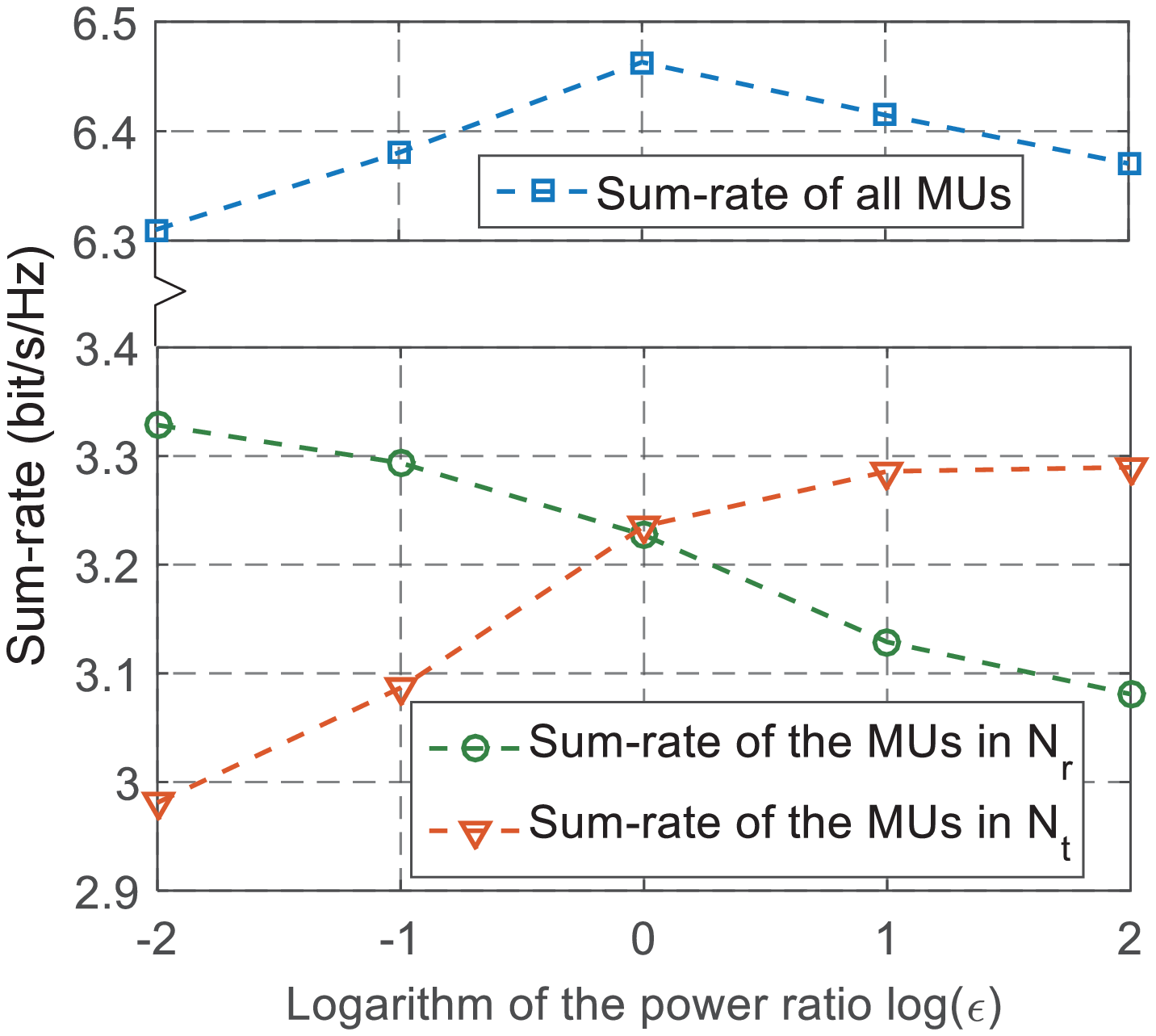}
\label{simulation3-1}
}%
\subfigure[MU distribution radius: 20 m.]{
\includegraphics[width=3in]{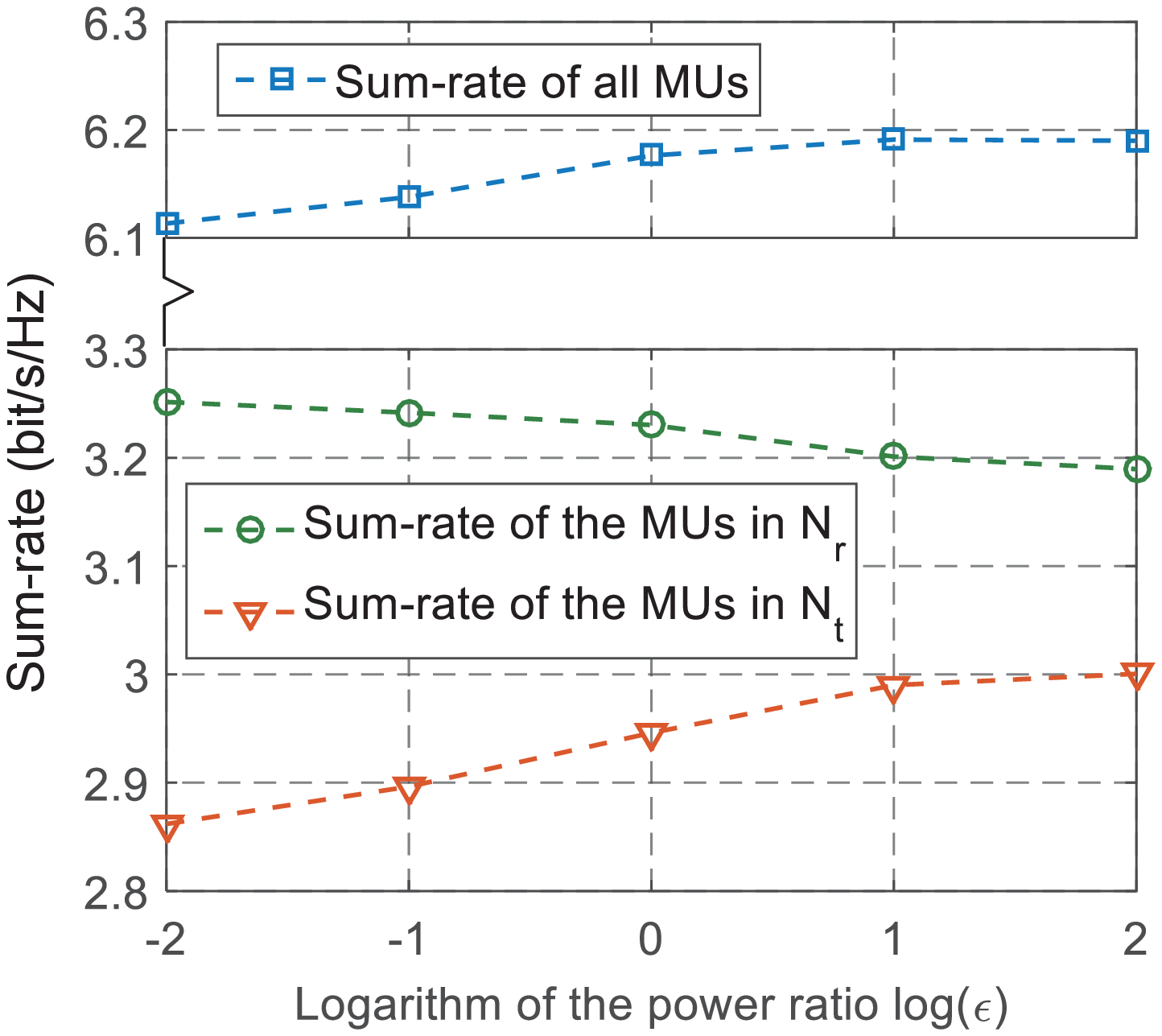}
\label{simulation3-2}
}%
\centering
\caption{Power ratio of the reflected and refracted signals ($\epsilon$) vs. the sum-rate.}
\label{simulation3}
\end{figure}
In Fig.~\ref{simulation3}, we analyze the relation between the average sum-rate and the power ratio of the reflected and refracted signals~$\epsilon$. In Fig.~\ref{simulation3-1}, the MUs are distributed around the IOS within a radius of 2 m, and the distance from the IOS to the MUs is much shorter than the distance from the SBS to the MUs. The average sum-rate of all the MUs is maximized for $\epsilon=1$, which is in agreement with Remark 3. In Fig.~\ref{simulation3-2}, the MUs are distributed around the IOS within a radius of 20 m. In this configuration, the MUs in $\mathcal{N}_r$ receive a higher power through the direct links when compared to the MUs in $\mathcal{N}_t$. The average sum-rate is maximized if $\epsilon>1$, which agrees with Proposition~6.

\begin{figure}[t]
    \centering
    \includegraphics[width=3.5in]{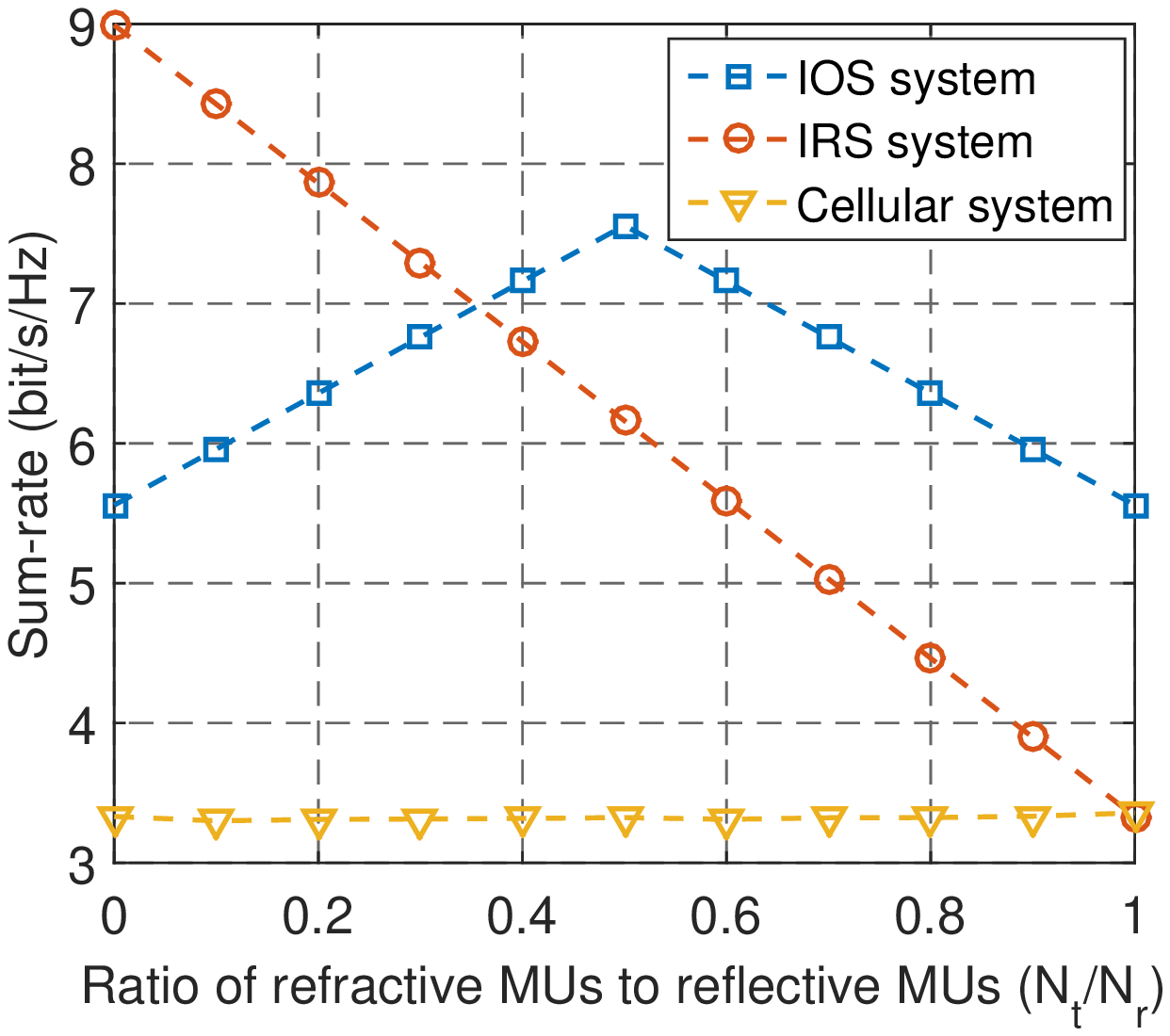}
    \caption{Ratio of MUs receiving refracted signals to all MUs ($N_t/N$) vs. sum-rate.}
    \label{simulation4}
\end{figure}

The impact of the distribution of the MU with respect to the location of the IOS is illustrated in Fig.~\ref{simulation4}. In the IOS-assisted communication system, the average sum-rate is maximized, in the considered setup, when the MUs are equally distributed on the two sides of the IOS (i.e., $N_r/N =0.5$ in the figure). When the MUs are mostly located on one side of the IOS, i.e., $N_t/N\rightarrow 0$ or $N_t/N \rightarrow 1$, the power of the signals re-emitted by the IOS on both sides of the surface cannot be optimized for all the MUs concurrently. In an IRS-assisted communication system, on the other hand, the average sum-rate decreases linearly as a function of $N_t/N$, since an RIS can assist the transmission of only the MUs in $\mathcal{N}_r$.

\begin{figure}[t]
    \centering
    \includegraphics[width=7in]{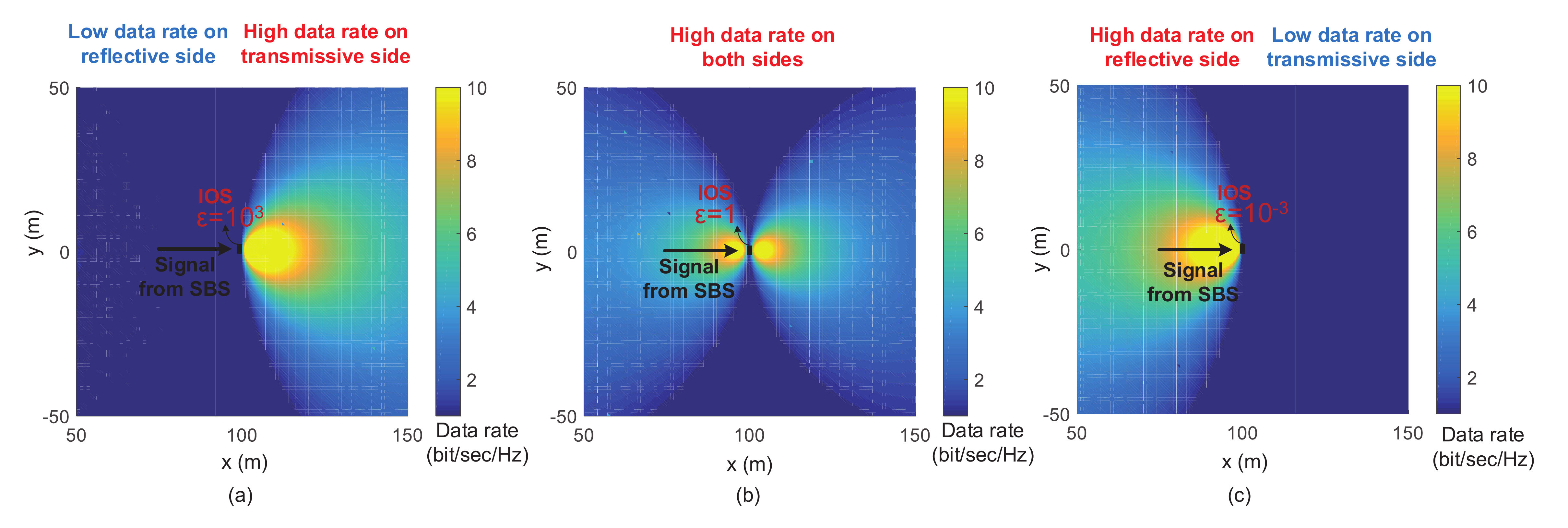}
    \caption{Simulation of the maximum data rate with different $\epsilon$. (a) $\epsilon=10^{3}$. (b) $\epsilon=1$. (c) $\epsilon=10^{-3}$.}
    \label{simulation5}
\end{figure}

In Fig.~\ref{simulation5}, we present the maximum sum-rate of a single MU at different locations, when the SBS is located on the left side of the surface. An IOS/IRS with $20\times 20$ elements is deployed vertically with respect to the SBS at the location (100, 0), and the SBS is deployed at (0,0). Three different values for the power ratio of the reflected and refracted signals $\epsilon$ are illustrated.

If $\epsilon=1$, the maximum sum-rate on both sides of the surface can be improved significantly. An omni-service extension is provided since the IOS is capable of enhancing the received power of the MUs on both sides of the surface. The MU can obtain a higher data rate when it is closer to the center of the IOS, where the reflective-refractive channel has a better quality. If $\epsilon=10^{3}$  and $\epsilon = 10^{-3}$, most of the power is re-emitted to the right-hand and left-hand side of the IOS, respectively. This is in agreement with Remark 1 and Remark 2.
\section{Conclusions}\label{Conclusion Sec}
In this paper, we have introduced the IOS, which is a reconfigurable surface capable of reflecting and refracting the impinging signals towards both sides of the surface concurrently. We have studied the utilization of the IOS in an indoor multi-user downlink communication system. The IOS is capable of enhancing the received signals of the MUs on both sides of the surface through an appropriate design of the IOS phase shifts. We have formulated a joint IOS analog beamforming and SBS digital beamforming optimization problem to maximize the sum-rate of the system, and have proposed an iterative algorithm to solve it efficiently. From the obtained numerical simulations and analysis, three main conclusions can be drawn:
\begin{enumerate}
\item The sum-rate of an IOS-assisted communication system is higher than that of an IRS-assisted communication system. An IOS enhances the data rate of the MUs located on both sides of the surface concurrently, which asymptotically double the achievable sum-rate and the service coverage.
\item The optimal power ratio of the reflected and refracted signals of the IOS is positively correlated with the ratio of the number of MUs on the two sides of the IOS, and is negatively correlated with the received power on the direct links from the SBS to the MUs.
\item The sum-rate of an IOS-assisted communication system increases with the geometric size of the surface and with the number of quantization bits of the phase shifts, but converges to a stable value when the IOS size and the quantization bits are sufficiently large.
\end{enumerate}

\begin{appendices}

\section{Proof of Proposition 3}
For a pair of MUs that are located symmetrically with respect to the IOS (e.g., the $i$th and $j$th MUs in Fig.~\ref{appendix D}(a)), the distance from each reconfigurable element to the two MUs is the same, i.e.,
\begin{equation}\label{Proposition3e1}
d_{m,i}=d_{m,j}, \forall m \in \mathcal{M}.
\end{equation}
According to~(\ref{Ko}), the power radiation pattern of the reflected signals $K^{D}(m)|_{r}$ and that of the refracted signals $K^{D}(m)|_{t}$ of the MUs that are located symmetrically with respect to the IOS satisfies the relation
\begin{equation}\label{Proposition3e2}
\frac{K^{D}(m)|_{r}}{K^{D}(m)|_{t}}=\epsilon.
\end{equation}
As shown in Fig.~\ref{appendix D}(a), in addition, the angles with respect to the normal to the surface satisfy the relation $\theta_i^D(m)+\theta_j^D(m)=\pi$, and then we have
\begin{equation}\label{Proposition3e3}
|\cos(\theta_i^D(m))|=|\cos(\theta_j^D(m))|.
\end{equation}
When substituting (\ref{Proposition3e1}), (\ref{Proposition3e2}), and (\ref{Proposition3e3}) into (\ref{signal gain}), we have $\frac{g_m(\xi^A(m),\xi_j^D(m),\psi_m)}{g_m(\xi^A(m),\xi_i^D(m),\psi_m)}=\epsilon$ for any IOS phase shifts $\bm{s}$. If we assume that the antenna gains of the MUs are the same, and substitute (\ref{signal gain}) into (\ref{main_channel}), we have $\frac{h_{j,k}^{m,LoS}}{h_{i,k}^{m,LoS}}=\epsilon,\ \forall m \in \mathcal{M}, k \in \mathcal{K}$, and for any IOS phase shifts $\bm{s}$. Under these assumptions, and if the small-scale fading is not considered, we evince that the optimal SBS digital beamforming and IOS analog beamforming of the $i$th and $j$th MUs coincide. This concludes the proof.

\section{Proof of Proposition 4}
The improvement of data rate in the presence of an IOS is determined by the sum of the channel gains of the $M$ SBS-IOS-MU links. Let us assume that the small scale fading can be ignored (long-term performance). From~(\ref{signal gain}) and (\ref{main_channel}), the LoS component of the SBS-IOS-MU link from the $k$th antenna of the SBS to the $i$th MU via the $m$th reconfigurable element of the IOS is
\begin{equation}\label{P4proof}
\begin{split}
h_{i,k}^{m,LoS}=\frac{\lambda K^{A}(m)K_i^{D}(m)\sqrt{G_k^{tx} G_m G_{i}^{rx}\delta_x\delta_y|\gamma_m|^2} \exp\Big(\frac{-j2\pi(d_{k,m}+d_{m,i})}{\lambda}-j \psi_{m}\Big)}{(4\pi)^{\frac{3}{2}}d_{k,m}^\alpha d_{m,i}^\alpha},
\end{split}
\end{equation}
The hardware-related parameters $G_k^{tx}$, $G_m$, $G_{i}^{rx}$, $\delta_x$, $\delta_y$, and $\gamma_m$ can be considered as constants in a given IOS-assisted communication system. The phase shift of the SBS-IOS-MU link $\frac{-j2\pi(d_{k,m}+d_{m,i})}{\lambda}-j \psi_{m}$ is optimized with the algorithm proposed in Section~\ref{IOS Phase Shift Design Sec}. Assuming that the locations of the SBS and IOS are given, the MU-related term that affects the channel gain of the LoS component is $\frac{K_i^{D}(m)}{d_{m,i}^\alpha},~\forall i \in \mathcal{N}, k \in \mathcal{K}, m \in \mathcal{M}$. Therefore, the impact of the $M$ reconfigurable elements can be expressed as
\begin{align}
	\mathcal{P}_i=\sum_{m=1}^M\frac{K_i^{D}(m)}{d_{m,i}^\alpha}=\left\{
	\begin{aligned}
		&\sum_{m=1}^M \frac{|\cos^3{\theta_i^{D}(m)}|}{(1+\epsilon)d_{m,i}^\alpha},~i \in \mathcal{N}_r,\\
		&\sum_{m=1}^M \frac{\epsilon|\cos^3{\theta_i^{D}(m)}|}{(1+\epsilon)d_{m,i}^\alpha},~i \in \mathcal{N}_t.
	\end{aligned}
    \right.
\end{align}
This concludes the proof.

\section{Proof of Proposition 5}
Assume that the signal-to-noise ratio~(SNR) of the direct link from the SBS to the $i$th MU is $\alpha_i$ and that the SNR of the link from the IOS to the $i$th MU is $\beta_i$. The rate of the $i$th MU is
\begin{equation}
R_i=\log_2(1+\alpha_i+\beta_i)\footnote{For simplicity, we do not consider the specific impact of the phase shift optimization, and assume that the SNRs of the SBS-to-MU link and the IOS-to-MU link can just be added directly.}.
\end{equation}
Therefore, the sum-rate of all the MUs in the set $\mathcal{N}$ is given by $\sum_{k \in \mathcal{N}_r} R_k+\sum_{j \in \mathcal{N}_r} R_j$.
Given the total power of the refracted and reflected signals, we aim to maximize the normalized sum-rate of the MUs, which is given as
\begin{equation}\label{delta R}
\sum_{i \in \mathcal{N}} R_i=\sum_{k \in \mathcal{N}_r}\log_2(1+\alpha_k+\frac{1}{1+\epsilon}\beta_k)+\sum_{j \in \mathcal{N}_t}\log_2(1+\alpha_j+\frac{\epsilon}{1+\epsilon}\beta_j).
\end{equation}
The first-order derivative of~(\ref{delta R}) with respect to $\epsilon$ is
\begin{equation}\label{appendixC equation}
\sum_{i \in \mathcal{N}}\frac{\emph{d} R_i}{\emph{d}\epsilon}=\frac{1}{\ln2}\left(\sum_{j \in \mathcal{N}_t}\frac{\beta_j/(1+\epsilon)^2}{1+\alpha_j+\frac{\beta_j\epsilon}{1+\epsilon}}-\sum_{k \in \mathcal{N}_r}\frac{\beta_k/(1+\epsilon)^2}{1+\alpha_k+\frac{\beta_k}{1+\epsilon}}\right).
\end{equation}
We denote the optimal value of $\epsilon$ that maximizes (\ref{delta R}) by $\epsilon^{opt}$. When the sum-rate is maximized, we have $\sum_{i \in \mathcal{N}}\frac{\emph{d}\Delta R_i}{\emph{d}\epsilon}|_{\epsilon=\epsilon^{opt}}=0$.

If the MU~$j'$ is added to the set $\mathcal{N}_t$, a positive value is added to~(\ref{appendixC equation}), and we have $\sum_{i \in \mathcal{N}}\frac{\emph{d}\Delta R_i}{\emph{d}\epsilon}|_{\epsilon=\epsilon^{opt}}$ $>0$. Thus, the sum-rate can be further improved by increasing $\epsilon$. On the contrary, if the MU~$k'$ is added to the set $\mathcal{N}_r$, a negative value is added to~(\ref{appendixC equation}). In this case a smaller value of $\epsilon$ improves the sum-rate of the system. This concludes the proof.

\section{Proof of Proposition 6}
We denote a pair of symmetrically-located MUs that receive the reflected signals and refracted signals by MU~$i$ and MU~$j$, respectively, as shown in Fig.~\ref{appendix D}(a). Based on the notation and formulas introduced in Appendix C, the normalized sum-rate of the MUs can be given as ($\beta_j = \beta_i$ since the MUs are located symmetrically with respect to the IOS)
\begin{equation}\label{delta R2}
R_i+R_j=\log_2\left(1+\alpha_i+\frac{1}{1+\epsilon}\beta_i\right)+\log_2\left(1+\alpha_j+\frac{\epsilon}{1+\epsilon}\beta_i\right).
\end{equation}
The derivative function of~(\ref{delta R2}) with respect to $\epsilon$ is
\begin{equation}
\frac{\emph{d}(R_i+R_j)}{\emph{d}\epsilon}= \frac{1}{\ln2}\frac{\beta_i^2(1-\epsilon)(1+\epsilon)+\beta_i(1+\epsilon)^2(\alpha_i-\alpha_j)}{(\epsilon\beta_i(1+\epsilon)+(1+\epsilon)^2)(\beta_i(1+\epsilon)+(1+\epsilon)^2)}.
\end{equation}
The optimal value of the power ratio of the reflected and refracted signals satisfies $\epsilon=\frac{\beta_i+\alpha_i-\alpha_j}{\beta_i-\alpha_i+\alpha_j}$. When $\alpha_i>\alpha_j$, we have $\epsilon>1$. Otherwise, we have $\epsilon<1$. This concludes the proof.
\section{Proof of Proposition 7}
Consider the two case studies illustrated in Fig.~\ref{appendix D}. In the IRS-assisted communication system, the total power that impinging upon the IRS is reflected towards the $i$th MU, and the downlink rate of the $i$th MU can be expressed as
\begin{equation}
R_i^{IRS}=\log_2(1+\alpha_i+\beta_i)+\log_2(1+\alpha_j).
\end{equation}
As far as the IOS-assisted communication system is concerned, on the other hand, the downlink rate can be expressed as
\begin{equation}\label{Propo 7 proof}
R_i^{IOS}=\log_2(1+\alpha_i+\frac{\beta_i}{\epsilon+1})+\log_2(1+\alpha_j+\frac{\epsilon\beta_i}{\epsilon+1}).
\end{equation}

Given the value of $\alpha_i$, $\beta_i$, and $\alpha_j$, $R_i^{IOS}$ is a function of $\epsilon$. The maximum value of $R_i^{IOS}$ can be obtained by analysing its derivate with respect to $\epsilon$.
\begin{subequations}\label{Propo 7 proof2}
\begin{align}
&\frac{\emph{d} R_i^{IOS}}{\emph{d} \epsilon}=\frac{\beta_i(\alpha_i-\alpha_j)}{(\epsilon+1)^2}+\frac{\beta_i^2(1-\epsilon)}{(\epsilon+1)^3},\label{Propo 7 proof2-1}\\
&\frac{\emph{d}^2 R_i^{IOS}}{\emph{d} \epsilon^2}=\frac{-2\beta_i\left((\alpha_i-\alpha_j-\beta_i^2)\epsilon+(\alpha_i-\alpha_j+2\beta_i^2)\right)}{(\epsilon+2)^4}.\label{Propo 7 proof2-2}
\end{align}
\end{subequations}
Without loss of generality, we assume that $\alpha_i\geq \alpha_j$. Since $\epsilon \in (0,1)$, we have $(\alpha_i-\alpha_j-\beta_i^2)\epsilon+(\alpha_i-\alpha_j+2\beta_i^2)\in (\alpha_i-\alpha_j+2\beta_i^2,2\alpha_i-2\alpha_j+\beta_i^2)$. Given that $\alpha_i-\alpha_j+2\beta_i^2>0$ and $2\alpha_i-2\alpha_j+\beta_i^2> 0$, the value of $(\alpha_i-\alpha_j-\beta_i^2)\epsilon+(\alpha_i-\alpha_j+2\beta_i^2)$ is larger than 0. Therefore, we have $\frac{\emph{d}^2 R_i^{IOS}}{\emph{d} \epsilon^2}<0$, which shows that $R_i^{IOS}$ is a concave function, and the maximum value is obtained with $\epsilon^{opt}=\frac{\alpha_i-\alpha_j+\beta_i}{\alpha_j-\alpha_i+\beta_i}$. When we substitute $\epsilon^{opt}$ into (\ref{Propo 7 proof}), the maximum rate is given by $R_i^{IOS,opt}=\log_2\left(1+\alpha_i+\alpha_j+\frac{\alpha_i\alpha_j}{2}+\frac{\alpha_i^2}{4}+\frac{\alpha_j^2}{4}+\beta_i+\frac{\beta_i(\alpha_i+\alpha_j)}{2}+\frac{\beta_i^2}{4}\right)$.
When comparing the value of $R_i^{IOS,opt}$ to two times of $R_i^{IRS}$, we have
\begin{equation}\label{Propo 7 proof3}
\begin{split}
2R_i^{IRS}=&\log_2(1+\alpha_i+\beta_i)^2(1+\alpha_j)^2\\
>&\log_2\left(1+\alpha_i+\alpha_j+\frac{\alpha_i\alpha_j}{2}+\frac{\alpha_i^2}{4}+\frac{\alpha_j^2}{4}+\beta_i+\frac{\beta_i(\alpha_i+\alpha_j)}{2}+\frac{\beta_i^2}{4}\right)
=R_i^{IOS,opt}.
\end{split}
\end{equation}
On the other hand, when $\alpha_i\rightarrow 0$, $\alpha_j\rightarrow 0$, and $\beta_i\rightarrow \infty$, we have
\begin{equation}\label{Propo 7 proof4}
\frac{R_i^{IOS,opt}}{R_i^{IRS}}\rightarrow 2.
\end{equation}
Combining (\ref{Propo 7 proof3}) and (\ref{Propo 7 proof4}), we conclude that the ratio of the downlink sum-rate of an IOS-assisted system and an IRS-assisted system is upper-bounded by two. This ends the proof.
\end{appendices}

\end{document}